\documentclass[aps,twocolumn,pre,floatfix]{revtex4-2}

\usepackage{xcolor,hyperref}
\hypersetup{
   colorlinks,
   linkcolor={blue!50!black},
   citecolor={blue!50!black},
   urlcolor={blue!80!black}
}

\usepackage{epsfig,amsmath}
\usepackage{bm}
\usepackage{soul,ulem}
\usepackage{hyperref}
\usepackage{footmisc}
\usepackage{wrapfig}
\DeclareMathAlphabet{\mathitbf}{OML}{cmm}{b}{it}

\newcommand{\tr}{\hbox{tr}}
\renewcommand{\=}{\!=\!}

\global\long\def\abs#1{\left|#1\right|}

\newcommand{\mat}[1]{\ensuremath{\boldsymbol{\mathrm{#1}}}}

\DeclareMathAlphabet\mathbfcal{OMS}{cmsy}{b}{n}
\begin{document}

\title{Breakdown of the classical rupture theory and\\ earthquake propagation in the ``forbidden'' super-Rayleigh range}
\author{Anna Pomyalov$^{1}$}
\author{Fabian Barras$^{2}$}
\author{Eran Bouchbinder$^{1}$}
\email{eran.bouchbinder@weizmann.ac.il}
\affiliation{$^{1}$Chemical and Biological Physics Department, Weizmann Institute of Science, Rehovot 7610001, Israel\\
$^{2}$The Njord Centre, Departments of Physics and Geosciences, University of Oslo, 0316 Oslo, Norway}

\begin{abstract}
Earthquakes propagating faster than the shear wave-speed are commonly thought to undergo a super-shear transition upon which they discontinuously jump from the sub-Rayleigh regime to the super-shear one. The super-Rayleigh regime, i.e., the range of propagation speeds between the Rayleigh and shear wave-speeds, is regarded as ``forbidden'' by the two-dimensional classical rupture theory. Here, we revisit the assumptions underlying the classical theory and develop a rupture theory that takes into account the dependence of the fault strength (frictional resistance) on the slip rate. The theory quantitatively agrees with numerical simulations nearly up to the Rayleigh wave-speed. Yet, very close to the latter, two-dimensional rupture solutions change their character due to frictional rate nonlinearity and rupture continuously propagates through the ``forbidden'' super-Rayleigh range into the super-shear regime, without a sharp super-shear transition. These results demonstrate that frictional rate dependence, generically observed in experiments, can have profound implications for fast earthquake propagation.
\end{abstract}

\maketitle

\section{I\lowercase{ntroduction}}

The propagation speed of earthquake rupture along a geological fault --- a prominent example of a frictional interface formed by two contacting bodies --- directly affects seismic energy radiation and consequently also the associated ground motion and seismic hazard~\cite{Scholz2002}. It is now established that a non-negligible fraction of the large earthquakes documented over the last few decades were super-shear, i.e., propagated at speeds exceeding the shear wave-speed $c_{\rm s}$~\cite{bao2022global,elbanna2025supershear}. Such super-shear ruptures are accompanied by the generation of Mach wavefronts, which potentially give rise to strong shaking over distances away from the fault that are significantly larger than those of sub-shear earthquakes~\cite{Rosakis2002,bizzarri2019mechanics,bernard2005shear,dunham2005near,bhat2007off}. Understanding the emergence of super-shear rupture is therefore a basic problem in earthquake physics.

A central aspect of this problem stems from the two-dimensional (2D) classical rupture theory~\cite{Freund1979,Freund1998,Broberg1999}, which predicts that shear rupture along homogeneous faults cannot propagate at speeds between the Rayleigh wave-speed $c_{_{\rm R}}$ (the speed of surface waves) and $c_{\rm s}$. Accepting this prediction, a major theoretical challenge has been to identify possible physical mechanisms to overcome the ``forbidden'' super-Rayleigh speeds range, allowing rupture to transition from the sub-Rayleigh regime to the super-shear one, i.e., to undergo a super-shear transition.

The earliest mechanism, and a prominent one, demonstrated that shear rupture along homogeneous 2D faults can nucleate secondary (`daughter') rupture ahead of the main (`mother') rupture edge~\cite{burridge1973admissible,andrews1976rupture,burridge1979stability,Gabriel2012,Svetlizky2016} --- depending on the propagation distance and the frictional stress conditions~\cite{dunham2007conditions} ---, which upon coalescence with the main rupture can give rise to a discontinuous jump from the sub-Rayleigh regime to the super-shear one. This discontinuous `mother-daughter' super-shear transition can be realized in three dimensions (3D) as well~\cite{dunham2007conditions}. Various other super-shear transition mechanisms have been explored, including fault heterogeneities  in both 2D and 3D (e.g., asperities or barriers ~\cite{fukuyama2002condition,dunham2003supershear,Liu2008,weng2015barrier,barras2017interplay}), fault steps overs~\cite{ryan2014dynamically,hu2016supershear,kehoe2020evidence}, fault roughness~\cite{Bruhat2016}, fault damage zone~\cite{Huang2016a}, geometric effects (e.g., free surface effects~\cite{kaneko2010supershear}), and genuine 3D effects~\cite{weng2020continuum}, among others. Super-shear rupture has also been documented in laboratory experiments, mainly on glassy polymers~\cite{rosakis1999cracks,xia2004laboratory,Ben-David2010a,Svetlizky2016,Kammer2018} but also on rocks~\cite{passelegue2013sub,xu2018strain}.

Here, we revisit the classical 2D rupture theory and its underlying assumptions, which lead to the prediction of a forbidden speeds range. One of these assumptions involves a residual frictional strength behind the rupture edge that is independent of the sliding rate, which is inconsistent with a broad range of observations that demonstrate the intrinsic and nonlinear rate dependence of frictional interfaces~\cite{Dieterich1979,Ruina1983,Shimamoto1986,tullis1986constitutive,blanpied1991fault,Persson1998,Marone1998a,Baumberger1999,Nakatani2001,Baumberger2006,Reches2010,Bar-Sinai2014,gou2024variation,Viesca2015,brantut2017fracture,bar2019spatiotemporal}. We develop a linearized rupture theory that relaxes this assumption and demonstrate its validity both well below the Rayleigh wave-speed $c_{_{\rm R}}$ and close to it using large-scale numerical simulations. Yet, the theory breaks down very close to $c_{_{\rm R}}$ due to frictional rate nonlinearity and numerical solutions reveal a continuous frictional rupture propagation through the forbidden super-Rayleigh speeds range into the super-shear regime, without a sharp super-shear transition. These findings demonstrate the possible limitations of the classical 2D rupture theory and highlight the importance of frictional rate dependence for fast earthquake propagation.

\vspace{-0.2cm}
\section{G\lowercase{eneralized rupture-edge singular fields and energy balance}}

Our first goal is to explicitly spell out the underlying assumptions of the classical 2D rupture theory and to explore their implications, which will also lead to our central theoretical development. To that aim, consider a fault/frictional interface formed by two large linear elastic bodies under the application of a constant compressive (normal) stress. That is, it is assumed that the two identical bodies follow the isotropic and homogeneous Hooke's law $(1+\nu)\mu\!\left[\nabla{\bm u}\!+\!(\nabla {\bm u})^{\mbox{\tiny T}}\right]\!=\!{\bm \sigma}\!-\!\nu({\bm I}\,\tr{\bm \sigma}\!-\!{\bm \sigma})$~\cite{Landau1986}, relating Cauchy's stress tensor ${\bm \sigma}(x,y,t)$ to the displacement vector ${\bm u}(x,y,t)$. Here, ${\bm I}$ is the identity tensor, $\nu$ is Poisson's ratio and $\mu$ is the shear modulus. The coordinates are chosen such that the interface lies along the $x$-axis, the direction normal to the interface is the $y$-axis (the interface is located at $y\!=\!0$) and $t$ is time. The Navier-Lam\'e linear elastodynamic equation~\cite{Landau1986} for ${\bm u}$ is obtained when Hooke's law is used inside the momentum equation $\nabla\cdot{\bm \sigma}\!=\!\rho\,\ddot{\bm u}$, where $\rho$ is the mass density and a superposed dot denotes a partial time derivative.

Consider then, in addition to the above-mentioned normal stress, denoted by $\sigma$, the application of a far-field shear stress $\tau_{\rm d}$ such that $\sigma_{xy}(x,y\=\pm\infty,t)\=\tau_{\rm d}$, which drives an interfacial shear rupture. The location of the rupture edge features a general time dependence $L(t)$ in the sub-shear range, i.e., $0\!\le\!c_{\rm r}(t)\!\le\!c_{\rm s}$, where $c_{\rm r}(t)\!\equiv\!\dot{L}(t)$. The edge separates a non-sliding interfacial state ahead of it from a sliding state behind it. That is, the slip (tangential displacement discontinuity) $\delta(x,t)\!\equiv\!2u_x(x,y\=0,t)$ satisfies $\delta(x\!\ge\!L(t),t)\=0$ and $\delta(x\!<\!L(t),t)\!>\!0$. The former implies that the fault/interface behaves elastically ahead of the rupture edge. Behind it, on the other hand, the shear stress is determined by the contact interaction between the sliding bodies, i.e., by the frictional strength $\tau$, such that $\sigma_{xy}(x\!<\!L(t),y\=0,t)\=\tau$. The problem formulation is completed once the dependence of $\tau$ on interfacial fields is explicitly indicated, i.e., when the interfacial constitutive relation is specified. We first proceed without specifying it.

A central assumption of the classical 2D rupture theory~\cite{Freund1998,Broberg1999} is that solutions to above-formulated problem, i.e., the resulting linear elastodynamic fields, admit a power-law series expansion in terms of a polar coordinate system $(r,\theta)$ co-moving with the rupture edge (such that its origin is located at $(x\!-\!L(t),y\=0)$ and $\theta\=0$ is the propagation direction)~\footnote{See Sect.~4.3 in~\cite{Freund1998}, Eq.~(4.3.2) therein in particular.}. Most importantly, it is assumed that the power-law series is dominated by a singular contribution to the displacement gradient proportional to $r^{\xi}$, with $-1\!<\!\xi\!<\!0$, sufficiently close to the edge~\footnote{In Eq.~(4.3.2) of~\cite{Freund1998}, displacement potentials are used such that the order of singularity therein is $p_{_{0}}\!=\!\xi+2$, with $1\!<\!p_{_{0}}\!<\!2$.}. That is, it is assumed that
\begin{equation}
    {\bm \nabla}{\bm u}(r,\theta,t) \sim r^{\xi} \qquad\hbox{with}\qquad -1 < \xi < 0 \ ,
\label{eq:edge_singularity}
\end{equation}
asymptotically in the small $r$ limit.

Equation~(\ref{eq:edge_singularity}) implies that the leading-order contribution to the shear stress ahead of the rupture edge takes the form $\sigma_{xy}(r,\theta\=0,t)\=K^{(\xi)}_{\rm II}r^{\xi}/\sqrt{2\pi}$, valid for any instantaneous, time-dependent propagation speed $c_{\rm r}(t)$~\cite{Freund1998}. Here, $K^{(\xi)}_{\rm II}$ quantifies the intensity of the singularity ($1/\sqrt{2\pi}$ is a convention), which cannot be obtained by an asymptotic analysis. In these terms, the slip rate $v(r,t)\=\dot{\delta}(r,t)$ and the shear stress behind the rupture edge take the form (see {\color{blue}{\em SI Appendix}})
\begin{eqnarray}
v(r,t)=\frac{K^{(\xi)}_{\rm II}}{\sqrt{2\pi}}r^{\xi}\sin(\pi\xi)\frac{2c_{\rm r}\,\alpha_{\rm s}}{\mu\,R(c_{\rm r})}\left(\alpha^2_{\rm s}-1\right) \ , \label{eq:vr}\\
\sigma_{xy}(r,\theta\!=\!\pm\pi,t)=\frac{K^{(\xi)}_{\rm II}}{\sqrt{2\pi}} r^{\xi}\cos(\xi\pi) \ .
\label{eq:stress_behind_edge}
\end{eqnarray}
Here, $\alpha_{\rm s,d}(c_{\rm r})\=\sqrt{1-c^2_{\rm r}/c^2_{\rm s,d}}$ and $R(c_{\rm r})\=4\alpha_{\rm d}\alpha_{\rm s}\!-\!(1+\alpha^2_{\rm s})^2$ are functions of $c_{\rm r}(t)$, where $c_{\rm d}$ is the dilatational wave-speed. The properties of the Rayleigh function $R(c_{\rm r})$, which vanishes at the Rayleigh wave-speed $c_{_{\rm R}}$, will be discussed later.

Equations~(\ref{eq:vr})-(\ref{eq:stress_behind_edge}) can be used to quantify rupture-edge energy balance. In this context, it is important to note that the singular fields discussed above cannot lead to real divergences, i.e., they are regularized over a small lengthscale, the so-called cohesive zone of size $\ell_{\rm c}$~\cite{Palmer1973,Freund1998,Broberg1999,berman2020dynamics}. That is, the near-edge singular fields in fact correspond to `intermediate asymptotics', valid on scales $r$ much smaller than the macroscopic scales characterizing the problem, but larger than the cohesive zone, i.e., $r\!\ge\!\ell_{\rm c}$. This implies, for example, that the slip rate continuously transitions from the no-sliding state ($v\=0$) ahead of the rupture edge to the sliding state ($v\!>\!0$) behind it, where it peaks at $r\!\sim\!\ell_{\rm c}$ and then follows Eq.~\eqref{eq:vr} at intermediate scales.

At these intermediate scales, the shear stress is expected to follow Eq.~\eqref{eq:stress_behind_edge} such that we can relate the frictional strength $\tau$ to the slip $\delta$. To proceed, we note that partial time derivatives in the problem can be related to spatial gradients according to $\partial_t\=c_{\rm r}\partial_r$, valid when $c_{\rm r}(t)$ does not vary rapidly. Consequently, we have $v(r,t)\!\simeq\!c_{\rm r}\partial\delta(r,t)/\partial{r}$, which upon using Eqs.~(\ref{eq:vr})-(\ref{eq:stress_behind_edge}) and the boundary condition $\sigma_{xy}(r,\theta\=\pm\pi,t)\=\tau$, leads to $\tau(\delta)\!\sim\!\delta^{\frac{\xi}{1+\xi}}$. The latter is valid for $r\!>\!\ell_{\rm c}$, i.e., for $\delta\!>\!\delta_{\rm c}$, where $\delta_{\rm c}$ is the slip accumulated when the rupture edge propagate a distance $\ell_{\rm c}$. To quantify the dissipation associated with frictional slip behind the edge, we define the ``breakdown energy'' $G_{\rm f}(\delta)\!\equiv\!\!\int_0^{\delta}[\tau(\delta')\!-\!\tau(\delta)]\,d\delta'$~\cite{Palmer1973,kanamori2000microscopic,Abercrombie2005,Tinti2005,Viesca2015,Nielsen2016}, from which we obtain (see {\color{blue}{\em SI Appendix}})
\begin{equation}
\label{eq:Gf_xiG_xi}
G_{\rm f}(\delta) = G_{\rm c}\,(\delta/\delta_{\rm c})^{^{\!\xi_{_{\rm G}}\!(\xi)}}\quad\quad\hbox{with}\quad\quad \xi_{_{\rm G}}\!(\xi)=\frac{1+2\xi}{1+\xi} \ ,
\end{equation}
for $\delta\!>\!\delta_{\rm c}$, where $G_{\rm c}\!\equiv\!G_{\rm f}(\delta_{\rm c})$ is the edge-localized dissipation.
\begin{figure*}[ht!]
\centering
\includegraphics[width=2.07\columnwidth]{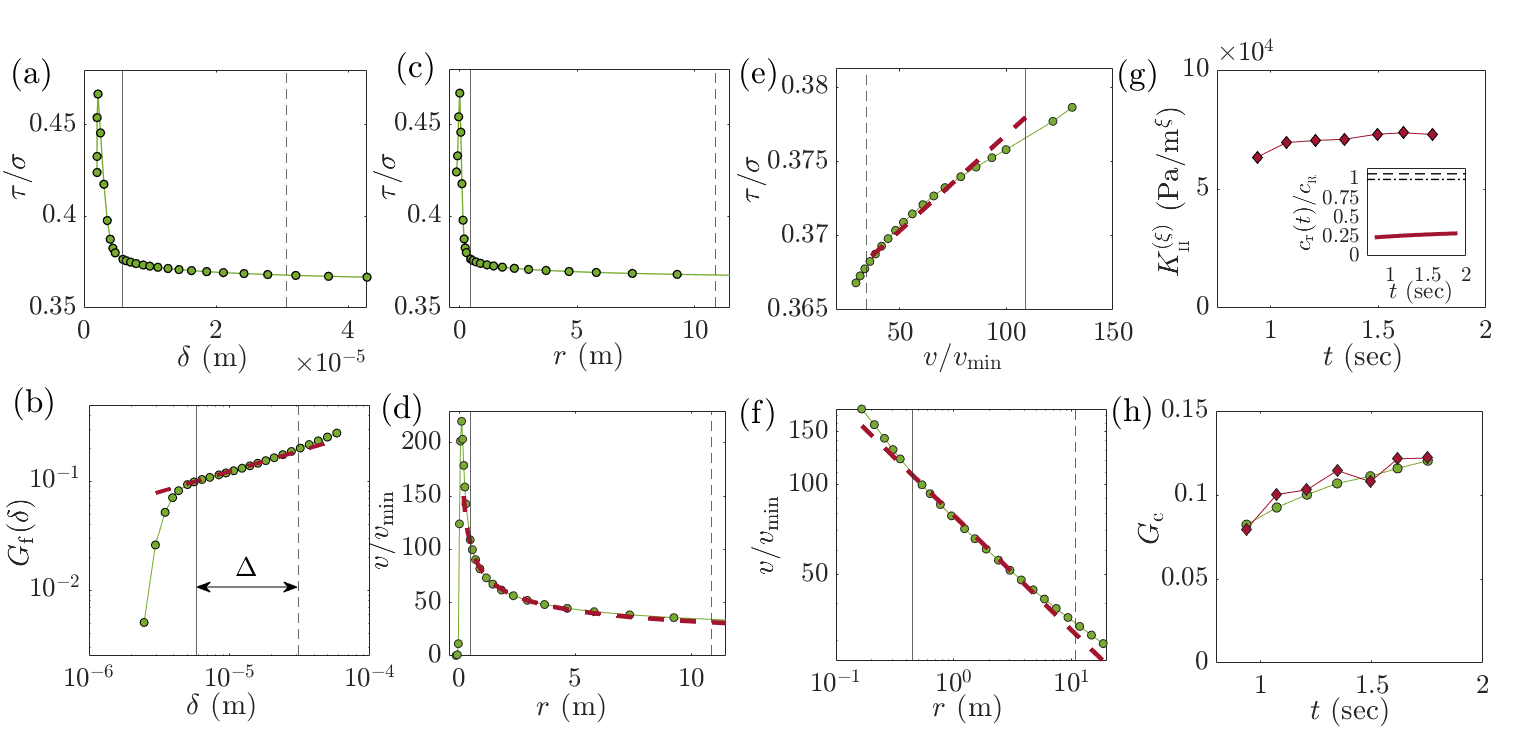}
\caption{{\bf Testing the unconventional singularity theory well below $c_{_{\rm R}}$}. A frictional rupture propagating at $c_{\rm r}(t)/c_{_{\rm R}}\!\simeq\!0.27$, obtained by a 2D boundary integral method computation, is used to test Eqs.~(\ref{eq:edge_singularity})-(\ref{eq:spectrum}). (a) $\tau(\delta)$ at a fixed position on the fault/interface during the passage of the rupture. The physical meaning of the vertical lines, here and elsewhere in the figure, will be explained below. Note that $\tau$ continues to weaken with slip $\delta$ even far beyond the large strength drop. (b) The corresponding breakdown energy $G_{\rm f}(\delta)$ (see text for definition). The superposed dashed line corresponds to a power-law predicted in Eq.~\eqref{eq:Gf_xiG_xi}, valid from $\delta_{\rm c}$ (vertical solid line) over a range $\Delta$ (up to the vertical dashed line). The double-logarithmic slope is $\xi_{_{\rm G}}$, from which $\xi\!\simeq\!-0.36$ is obtained using Eq.~\eqref{eq:Gf_xiG_xi}. (c) $\tau(r)$ behind the rupture edge, cf.~the prediction in Eq.~\eqref{eq:stress_behind_edge}, where $r\!=\!|x_{\rm p}\!-\!x|$ and $x_{\rm p}$ is the peak shear stress position. The cohesive zone of size $\ell_{\rm c}$, appearing in Eq.~\eqref{eq:G_unconventional}, is the distance between $x_{\rm p}$ and the vertical solid line (defined above). (d) The corresponding slip rate field $v(r)$, cf.~the prediction in Eq.~\eqref{eq:vr}. The superposed dashed line is discussed in panel (f). (e) $\tau(v)$, where the superposed dashed line is a linear fit between the vertical lines, corresponding to Eq.~\eqref{eq:effective_linearization}. It results in $\eta_{\rm eff}$, which upon solving Eq.~\eqref{eq:spectrum}, yields $\xi\!\simeq\!-0.36$, in excellent agreement with the value independently obtained in panel (b). (f) The same as panel (d), but on a double-logarithmic scale. The superposed dashed line is the prediction in Eq.~\eqref{eq:vr} with $\xi$ predicted in panels (b) and (e). The same line is superposed in panel (d), both also determining $K^{(\xi)}_{\rm II}$, the generalized stress-intensity-factor. (g) $K^{(\xi)}_{\rm II}$ vs.~$t$ and (inset) the corresponding $c_{\rm r}(t)$. (h) Testing Eq.~\eqref{eq:G_unconventional}, where the RHS $G_{\rm c}(t)$ (diamonds) is obtained using Eq.~\eqref{eq:Gf_xiG_xi} as in panel (b) according to $G_{\rm c}\!=\!G_{\rm f}(\delta_{\rm c})$, and the LHS (circles) using the physical parameters extracted above. See text and {\color{blue}{\em SI Appendix}} for additional details and extracted parameter values.}
\label{fig:unconventional_singularity_low}
\end{figure*}

Finally, $G_{\rm c}$ --- that quantifies the near-edge dissipation over the cohesive zone scale $\ell_{\rm c}$ --- is balanced by the flux of elastic energy into the edge region, which is quantified by the energy release rate $G$~\cite{Freund1998}. The latter can be computed using the above-derived singular fields, leading to
\begin{equation}
\label{eq:G_unconventional}
    G = \frac{(1-\nu)}{2\mu}A_{\rm II}(c_{\rm r},\xi)\left[K^{(\xi)}_{II}\right]^2 \ell_{\rm c}^{^{2\xi+1}} = G_{\rm c} \ ,
\end{equation}
where $A_{\rm II}(c_{\rm r},\xi)$ is given in the {\color{blue}{\em SI Appendix}}. Equation~(\ref{eq:G_unconventional}) constitutes rupture-edge energy balance.

Equations~(\ref{eq:edge_singularity})-(\ref{eq:G_unconventional}) are valid for any $\xi$ in the range $-1\!<\!\xi\!<\!0$. It is crucial to stress that Eqs.~(\ref{eq:edge_singularity})-(\ref{eq:G_unconventional}) directly follow from the assumptions of the classical 2D rupture theory, i.e., bulk linear elastodynamics and a power-law series solution dominated by a singular contribution, with no additional assumptions, prior to the selection of the singularity order $\xi$. The latter and its implications next are discussed next.

\vspace{-0.2cm}
\section{T\lowercase{he selection of the singularity order and the forbidden speeds range}}

The singularity order $\xi$ is selected by the boundary condition $\sigma_{xy}(r,\theta\=\pm\pi,t)\=\tau$ once the dependence of the frictional strength $\tau$ on other interfacial fields --- i.e., the interfacial/fault constitutive relation (the `friction law') --- is specified. The friction law involves strongly non-equilibrium, nonlinear, and dissipative physics and may vary from fault to fault. Our goal here is not to consider a specific and very detailed friction law, but rather to identify generic and minimal properties of the frictional strength and explore their implications.

A very large body of experimental evidence --- accumulated over several decades --- showed that the frictional strength depends on the slip rate $v$ and on the state of the fault~\cite{Dieterich1979,Ruina1983,Shimamoto1986,tullis1986constitutive,blanpied1991fault,Persson1998,Marone1998a,Baumberger1999,Nakatani2001,Baumberger2006,Reches2010,Bar-Sinai2014,gou2024variation,Viesca2015,brantut2017fracture,bar2019spatiotemporal}, expressed as $\tau(v,\ldots)$. Here, the ellipsis stands for interfacial state fields (e.g., roughness-induced contact area, gouge layer, fluid pore pressure, temperature etc.) that follow their own evolution equations. Generically, $\tau(v,\ldots)$ is {\em nonlinear}. To make theoretical progress, we focus on scales behind the rupture edge larger than $\ell_{\rm c}$ (i.e., on $\delta\!>\!\delta_{\rm c}$), where the frictional strength is expected to be predominantly slaved to the slip rate, i.e., to be given by $\tau(v)$. We then expand the latter as~\cite{Brener2021unconventional,Brener2021JMPS}
\begin{equation}
\tau(v) \simeq \tau_0 + \eta_{\rm eff}\,v + {\cal O}(v^2)\ ,
\label{eq:effective_linearization}
\end{equation}
over some range of slip rates, where $\eta_{\rm eff}$ is an effective viscous-friction coefficient.

Equation~(\ref{eq:effective_linearization}) allows to discuss another crucial assumption of the classical rupture theory. The latter assumes no rate dependence, implying that $\tau(v)\!\simeq\!\tau_0$, where $\tau_0$ is a constant residual stress. Consequently, the boundary condition $\sigma_{xy}(r,\theta\=\pm\pi,t)\=\tau_0$ implies that the singular contribution to the shear stress behind the rupture edge in Eq.~\eqref{eq:stress_behind_edge} must vanish, i.e., that $\xi\=-1/2$, which is nothing but the conventional square root singularity of the classical theory~\cite{Freund1998}. Substituting $\xi\=-1/2$ in Eqs.~(\ref{eq:edge_singularity})-(\ref{eq:G_unconventional}), all classical results are recovered~\cite{Freund1998}. Most importantly, $\xi_{_{\rm G}}\!(\xi\=-1/2)\=0$ in Eq.~\eqref{eq:Gf_xiG_xi}, which implies $G_{\rm f}(\delta)\=G_{\rm c}$ for $\delta\!>\!\delta_{\rm c}$, $K_{\rm II}\!\equiv\!K^{(-1/2)}_{\rm II}$ is the classical stress-intensity-factor, the dependence of $G$ on $\ell_{\rm c}$ in Eq.~\eqref{eq:G_unconventional} is eliminated and $A_{\rm II}(c_{\rm r},\xi\=-1/2)\=(c_{\rm r}/c_{\rm s})^2\alpha_s/\left[(1-\nu)R(c_{\rm r})\right]$ therein. The latter is the origin of the ``forbidden'' super-Rayleigh speeds range in the classical theory, as explained next.

The Rayleigh function $R(c_{\rm r})\=4\alpha_{\rm d}\alpha_{\rm s}\!-\!(1+\alpha^2_{\rm s})^2$ is positive in the sub-Rayleigh regime, $0\!\le\!c_{\rm r}\!<\!c_{_{\rm R}}$, vanishes at the Rayleigh wave-speed, $c_{\rm r}\=c_{_{\rm R}}$, and becomes negative in the super-Rayleigh regime, $c_{_{\rm R}}\!<\!c_{\rm r}\!<\!c_{\rm s}$. Consequently, $A_{\rm II}(c_{\rm r},\xi\=-1/2)\!\sim\!1/R(c_{\rm r})$ diverges as $c_{\rm r}\!\to\!c_{_{\rm R}}$ and becomes negative in $c_{_{\rm R}}\!<\!c_{\rm r}\!<\!c_{\rm s}$, implying that the rupture-edge energy balance $G\=G_{\rm c}$ cannot be satisfied (note that $G_{\rm c}\!>\!0$ for all speeds). Moreover, the slip rate in Eq.~\eqref{eq:vr} with $\xi\=-1/2$ is predicted to feature the same divergence and sign change, which appear to violate physical expectation. Consequently, the classical theory excludes rupture propagation at super-Rayleigh speeds~\cite{Freund1998}.

The assumption about the lack of frictional rate dependence, i.e., $\tau(v)\!\simeq\!\tau_0$, is clearly inconsistent with a broad range of observations~\cite{cocco2023fracture}. Our next goal is to relax this assumption and thoroughly explore its implications for frictional rupture (earthquake) propagation, especially at high speeds. Therefore, we use the leading-order rate dependent term in Eq.~\eqref{eq:effective_linearization}, which preserves linearity and allows analytical progress, inside the frictional boundary condition $\sigma_{xy}(r,\theta\=\pm\pi,t)\=\tau(v)$. Together with Eqs.~(\ref{eq:vr})-(\ref{eq:stress_behind_edge}), we obtain
\begin{equation}
\cot(\pi\xi) \simeq - \frac{\eta_{\rm eff}\,c_{\rm s}}{\mu}\frac{2(c_{\rm r}/c_{\rm s})^3 \alpha_{\rm s}}{R(c_{\rm r})} \ ,
\label{eq:spectrum}
\end{equation}
which selects the singularity order $\xi$. It is important to note that Eq.~\eqref{eq:spectrum}, whose properties will be further discussed below, predicts that $\xi$ is non-universal, depending on both the rupture speed $c_{\rm r}$ and the friction law, through $\eta_{\rm eff}(c_{\rm r})$.

Equations~(\ref{eq:edge_singularity})-(\ref{eq:spectrum}) constitute the unconventional singularities theory of frictional rupture. It builds on, and expands, the recent developments of~\cite{Brener2021unconventional,Brener2021JMPS}. Unconventional singularity orders, $\xi\!\ne\!-1/2$, have emerged in the framework of various rate-dependent constitutive laws, including viscous-friction~\cite{Ida1974slow,Brener2002,Brener2005,Brener2021unconventional,Brener2021JMPS}, thermal pressurization~\cite{Viesca2015}, flash heating~\cite{brantut2017fracture}, as well as in hydraulic fracture~\cite{Desroches1994,garagash2000tip,detournay2016mechanics}. Moreover, a non-constant breakdown energy $G_{\rm f}(\delta)$ for $\delta\!>\!\delta_{\rm c}$, i.e., Eq.~\eqref{eq:Gf_xiG_xi} with $\xi_{_{\rm G}}\!(\xi)\!>\!0$, has been extensively discussed in recent literature, e.g.,~\cite{Viesca2015,brantut2017fracture,Barras2020,Brener2021unconventional,cocco2023fracture,paglialunga2022scale,paglialunga2024frictional,fryer2024effect}, though not in the context of super-Rayleigh rupture propagation and the super-shear transition. Equations~(\ref{eq:edge_singularity})-(\ref{eq:spectrum}), some of which already received recent experimental support~\cite{paglialunga2022scale,paglialunga2024frictional}, have not been systematically and quantitatively tested so far against frictional rate-dependent elastodynamic simulations of in-plane shear rupture, which is our next goal.

\vspace{-0.2cm}
\section{T\lowercase{esting the unconventional singularities theory}}

To test the theory, to be abbreviated hereafter as UST, we performed large-scale 2D boundary integral method simulations employing a nonlinear, rate-and-state friction law~\cite{roch2022cracklet}. The details of the friction law are given in the {\color{blue}{\em SI Appendix}}, where its single most important property in the present context is that its steady sliding frictional strength $\tau_{\rm ss}(v)$ depends nonlinearly on the slip rate $v$ and features an $N$ shape, observed in numerous experiments~\cite{Bar-Sinai2014,gou2024variation}. The high slip rates rate-strengthening branch of this $N$-shaped curve, above its local minimum at $(v_{\rm min}, \tau_{\rm min})$, is expected to be relevant at the tail of a propagating earthquake, potentially making Eq.~\eqref{eq:effective_linearization} with $\eta_{\rm eff}\!>\!0$ relevant.

We generate rupture by the earthquake nucleation procedure explained in {\color{blue}{\em SI Appendix}} under various background stress (prestress) levels $\tau_{\rm d}$. For a fixed earthquake nucleation procedure, friction parameters and fault size, $c_{\rm r}(t)$ is mainly controlled by $\tau_{\rm d}$. We tracked it, along with the spatiotemporal evolution of the slip rate $v(x,t)$ and of the fault strength $\tau(x,t)$ fields. These computations allow to quantitatively test Eqs.~(\ref{eq:edge_singularity})-(\ref{eq:spectrum}). It is crucial to stress that the latter offer very stringent predictions; for example, $\xi$ can be obtained from $G_{\rm f}(\delta)$ of Eq.~\eqref{eq:Gf_xiG_xi} and independently from Eq.~\eqref{eq:spectrum}, once $\eta_{\rm eff}$ is extracted.

In Fig.~\ref{fig:unconventional_singularity_low}, we present a very comprehensive test of Eqs.~(\ref{eq:edge_singularity})-(\ref{eq:spectrum}) for a slowly accelerating rupture propagating well below $c_{_{\rm R}}$, here $c_{\rm r}(t)/c_{_{\rm R}}\!\simeq\!0.27$. The results demonstrate (see figure caption for details) that the theory is in very good quantitative agreement with all the numerical observations with an unconventional singularity order $\xi\!\simeq\!-0.36$, independently and self-consistently obtained from both extracting $\xi_{_{\rm G}}\!(\xi)$ in Eq.~\eqref{eq:Gf_xiG_xi} and from Eq.~\eqref{eq:spectrum}, involving the measured $\eta_{\rm eff}$. The results also imply that the classical 2D rupture theory with $\xi\=-1/2$ fails to quantitatively account for the numerical data.

\begin{figure}[ht!]
\centering
\includegraphics[width=0.9\columnwidth]{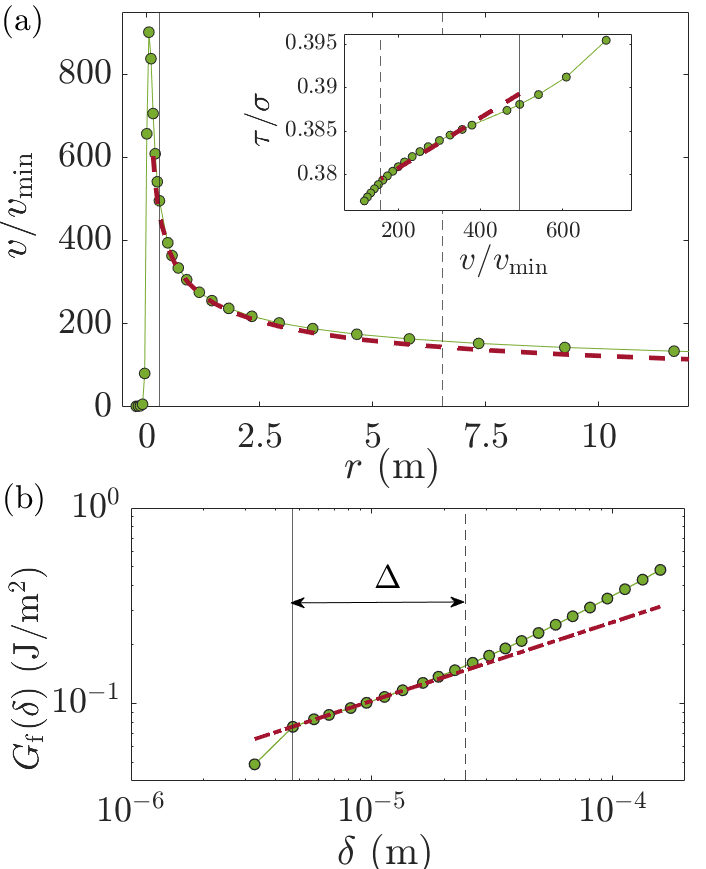}
\caption{{\bf Testing the unconventional singularities theory close to $c_{_{\rm R}}$}. Similarly to Fig.~\ref{fig:unconventional_singularity_low}, but for $c_{\rm r}(t)\!\simeq\!0.93c_{_{\rm R}}$. (a) As in Fig.~\ref{fig:unconventional_singularity_low}d. (inset) As in Fig.~\ref{fig:unconventional_singularity_low}e. (b) As in Fig.~\ref{fig:unconventional_singularity_low}b. The obtained unconventional singularity order is $\xi\!=\!-(0.34 \pm 0.03)$, where the upper value corresponds to panel (b) and the lower to the inset of panel (a), see text and {\color{blue}{\em SI Appendix}} for details and all extracted parameter values.}
\label{fig:unconventional_singularity_high}
\end{figure}

Next, we consider significantly larger propagation speeds. Specifically, we increased the background stress $\tau_{\rm d}$ and focus on a time snapshot for which $c_{\rm r}(t)/c_{_{\rm R}}\!\simeq\!0.93$, i.e., relatively close to $c_{_{\rm R}}$. The analysis is presented in Fig.~\ref{fig:unconventional_singularity_high} and demonstrates yet again good quantitative agreement between the UST and the numerical observations with $\xi\=-(0.34 \pm 0.03)$, where the small uncertainty reflects a small difference in the predictions of Eq.~\eqref{eq:Gf_xiG_xi} and Eq.~\eqref{eq:spectrum}. These results further highlight the breakdown of the classical 2D rupture theory with $\xi\=-1/2$ in our numerical data and the success of the UST up to high propagation speeds. They also raise the following question: what does the UST predict for rupture propagation very close to $c_{_{\rm R}}$ and in the super-Rayleigh range?

\vspace{-0.2cm}
\section{A\lowercase{pproaching $c_{_{\uppercase{\rm R}}}$ and super-\uppercase{R}ayleigh propagation}}

To start addressing the central question posed above, we first analyze the prediction for $\xi$ in Eq.~\eqref{eq:spectrum} and its $c_{\rm r}$ dependence as $c_{_{\rm R}}$ is approached. Since $R(c_{\rm r}\!\to\!c_{_{\rm R}})\!\to\!0$, the right-hand-side (RHS) of Eq.~\eqref{eq:spectrum} diverges in this limit, implying that the left-hand-side (LHS) $\cot[\pi\xi(c_{\rm r})]\!\sim\!1/\sin[\pi\xi(c_{\rm r})]$ also diverges, while $\sin[\pi\xi(c_{\rm r})]/R(c_{\rm r})$ remains finite. Consequently, Eq.~\eqref{eq:spectrum} predicts that $\xi\!\to\!0$ for $c_{\rm r}\!\to\!c_{_{\rm R}}$, and that the slip rate in Eq.~\eqref{eq:vr} --- that is proportional to $\sin[\pi\xi(c_{\rm r})]/R(c_{\rm r})$ --- does not experience any divergence in this limit, in sharp contrast to the corresponding prediction of the classical theory with $\xi\=-1/2$. These predictions assume that $\eta_{\rm eff}(c_{\rm r})$ reveals no special behavior in the vicinity of $c_{_{\rm R}}$, which is expected on general grounds and verified ({\color{blue}{\em SI Appendix}}).

A vanishing singularity order, $\xi\!\to\!0$, amounts to rupture propagation without stress amplification. This can be excluded on physical grounds since a stress amplification relative to the prestress in the close vicinity of the rupture edge is necessary for interfacial failure and for the propagation of frictional rupture. Consequently, the UST appears to signal its own breakdown in the $c_{\rm r}\!\to\!c_{_{\rm R}}$ limit. Since Eq.~\eqref{eq:spectrum} is equivalent to the frictional boundary condition $\sigma_{xy}(r,\theta\=\pm\pi,t)\=\tau(v)$ under the linearization of $\tau(v)$ in Eq.~\eqref{eq:effective_linearization}, this breakdown suggests that nonlinearity in $\tau(v)$ becomes increasingly important in the $c_{\rm r}\!\to\!c_{_{\rm R}}$ limit. In the presence of such nonlinearity, the power-law series assumption of the classical rupture theory cannot hold and $\xi$ can no longer be selected.

A direct and transparent way to probe the emergence of frictional rate nonlinearity in $\tau(v)$ in the $c_{\rm r}\!\to\!c_{_{\rm R}}$ limit is through the behavior of $G_{\rm f}(\delta)$ for $\delta\!>\!\delta_{\rm c}$. On the one hand, the UST in Eq.~\eqref{eq:Gf_xiG_xi} predicts that $G_{\rm f}(\delta)$ varies as a power-law with $\xi_{_{\rm G}}\!(\xi\!\to\!0)\!\to\!1$ in this limit. On the other hand, the above considerations suggest that $G_{\rm f}(\delta)$ will not reveal a clean power-law with $\xi_{_{\rm G}}\!\to\!1$. To quantitatively distinguish between the two, we quantify the power-law scaling range of $G_{\rm f}(\delta)$, denoted by $\Delta$ in Fig.~\ref{fig:unconventional_singularity_low}b and Fig.~\ref{fig:unconventional_singularity_high}b, where the emergence of frictional rate nonlinearity in $\tau(v)$ would be manifested in a significant reduction in $\Delta$.

\begin{figure}[ht!]
\centering
\includegraphics[width=1\columnwidth]{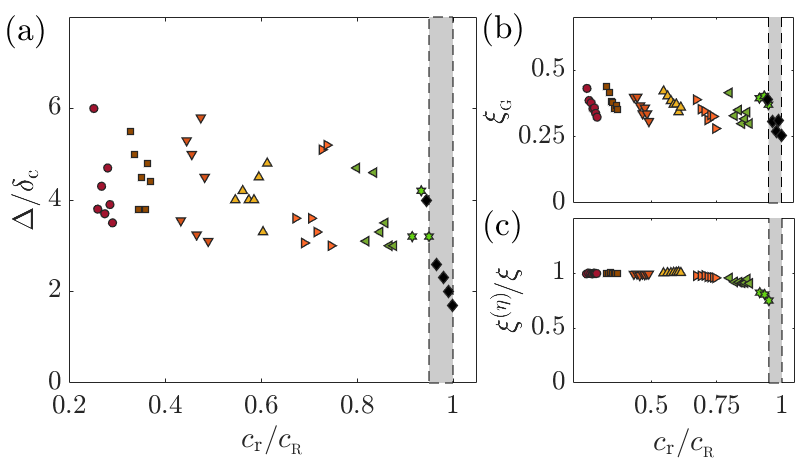}
\caption{{\bf The breakdown of the unconventional singularities theory approaching $c_{_{\rm R}}$}. (a) $\Delta(c_{\rm r})/\delta_{\rm c}$, quantified as in Fig.~\ref{fig:unconventional_singularity_low}b and~\ref{fig:unconventional_singularity_high}b. Each color/symbol corresponds to a different value of the prestress $\tau_{\rm d}$, increasing from left to right, see {\color{blue}{\em SI Appendix}}. The shaded region starts at $c_{\rm r}\!\simeq\!0.93c_{_{\rm R}}$, corresponding to the rupture that is analyzed in Fig.~\ref{fig:unconventional_singularity_high} and shown therein to be quantitatively consistent with the UST, see text. It is observed that $\Delta(c_{\rm r})$ slowly decreases in the range $0.27c_{_{\rm R}}\!<\!c_{\rm r}\!<\!0.93c_{_{\rm R}}$ (with some variation within each $\tau_{\rm d}$ dataset), but significantly decreases for $c_{\rm r}\!>\!0.93c_{_{\rm R}}$, corresponding to the emergence of significant frictional rate nonlinearity as $c_{\rm r}\!\to\!c_{_{\rm R}}$. It marks the breakdown of the UST as $c_{_{\rm R}}$ is approached. (b) The power-law scaling exponent $\xi_{_{\rm G}\!}(c_{\rm r})$ corresponding to panel (a). Values in the shaded region ($0.93c_{_{\rm R}}\!<\!c_{\rm r}\!<\!c_{_{\rm R}}$) are reported, despite the narrow scaling range in the shaded region shown in panel (a), to highlight the qualitative difference compared to $\xi_{_{\rm G}\!}(c_{\rm r}\!\to\!c_{_{\rm R}})\!\to\!1$ (predicted by Eqs.~(\ref{eq:Gf_xiG_xi}) and~(\ref{eq:spectrum})). (c)  The ratio of $\xi^{(\eta)}$, predicted by Eq.~\eqref{eq:spectrum}, and $\xi$, obtained by inverting $\xi_{_{\rm G}\!}(\xi)$ of Eq.~\eqref{eq:Gf_xiG_xi}, vs.~$c_{\rm r}\!<\!0.93c_{_{\rm R}}$. $\xi^{(\eta)\!}\!/\xi$ is close to unity across the entire range considered, with a small decline as $c_{\rm r}\!\to\!0.93c_{_{\rm R}}$, see text.}
\label{fig:breakdown_and_transition}
\end{figure}

In Fig.~\ref{fig:breakdown_and_transition}a, we plot $\Delta(c_{\rm r})/\delta_{\rm c}$ that is quantified exactly as in Fig.~\ref{fig:unconventional_singularity_low}b and Fig.~\ref{fig:unconventional_singularity_high}b, i.e., by estimating the $\delta$ range over which $G_{\rm f}(\delta)$ follows a straight line for $\delta\!>\!\delta_{\rm c}$ in a double-logarithmic representation (see {\color{blue}{\em SI Appendix}}). It is observed that $\Delta(c_{\rm r})/\delta_{\rm c}$ slowly decreases in the range $0.27c_{_{\rm R}}\!<\!c_{\rm r}\!<\!0.93c_{_{\rm R}}$ (with some variation within each dataset, see caption), where we already established the validity of the UST, but rather strongly decreases for $0.93c_{_{\rm R}}\!<\!c_{\rm r}\!<\!c_{_{\rm R}}$ (shaded region).

While the values of $\Delta(c_{\rm r})/\delta_{\rm c}$ in the shaded region become too small for a power-law to be meaningfully defined, we nevertheless plot $\xi_{_{\rm G}\!}(c_{\rm r})$ (the slope of the line in the double-logarithmic representation of $G_{\rm f}(\delta)$ that is used to determine $\Delta(c_{\rm r})$) in Fig.~\ref{fig:breakdown_and_transition}b. Recall, that Eq.~\eqref{eq:spectrum} together with Eq.~\eqref{eq:Gf_xiG_xi} predict $\xi_{_{\rm G}\!}(c_{\rm r}\!\to\!c_{_{\rm R}})\!\to\!1$. Yet, Fig.~\ref{fig:breakdown_and_transition}b reveals no sign of such a behavior as $c_{_{\rm R}}$ is approached. Taken together, Figs.~\ref{fig:breakdown_and_transition}a-b strongly indicate that the UST fails close to $c_{_{\rm R}}$, most likely due to frictional rate nonlinearity. To reiterate and highlight the success of the theory not very close to $c_{_{\rm R}}$, already demonstrated in Figs.~\ref{fig:unconventional_singularity_low}-\ref{fig:unconventional_singularity_high}, we denote by $\xi^{(\eta)}$ the prediction in Eq.~\eqref{eq:spectrum} --- to distinguish it from $\xi$ that is obtained by inverting $\xi_{_{\rm G}\!}(\xi)$ of Eq.~\eqref{eq:Gf_xiG_xi} --- and plot their ratio in Fig.~\ref{fig:breakdown_and_transition}c for $0.27c_{_{\rm R}}\!<\!c_{\rm r}\!<\!0.93c_{_{\rm R}}$. It is observed that $\xi^{(\eta)\!}/\xi$ is close to unity across the entire range considered, with a small decline as $c_{\rm r}\!\simeq\!0.93c_{_{\rm R}}$ is approached, already alluded to above in relation to Fig.~\ref{fig:unconventional_singularity_high}.

The results in Fig.~\ref{fig:breakdown_and_transition} clearly indicate that the UST is valid over a broad range of rupture propagation speeds, but fails as $c_{\rm r}\!\to\!c_{_{\rm R}}$ due to an increasing importance of frictional rate nonlinearity. The most remarkable implication of these results is that there is no reason to expect the Rayleigh function $R(c_{\rm r})$, and consequently its zero at $c_{\rm r}\=c_{_{\rm R}}$, to appear in the nonlinear theory close to the $c_{_{\rm R}}$. In other words, the Rayleigh wave-speed $c_{_{\rm R}}$ is not expected to be a `special propagation speed' anymore and likewise one expects that $G\!>\!0$ in the super-Rayleigh range of speeds $c_{_{\rm R}}\!<\!c_{\rm r}\!<\!c_{\rm s}$. If so, frictional rupture propagation in the ``forbidden'' super-Rayleigh range should be possible. Indeed, this is demonstrated in Fig.~\ref{fig:super_Rayleigh}, where rupture smoothly and continuously propagates into the ``forbidden'' super-Rayleigh range for various prestress levels $\tau_{\rm d}$ (the inset presents $v(r)$ as $c_{_{\rm R}}$ is crossed for one $\tau_{\rm d}$ value, see caption). This is a central finding of this study.

\vspace{-0.2cm}
\section{A \lowercase{continuous super-shear crossover}}

The results discussed so far demonstrated that the generic rate dependence of the frictional strength gives rise to the breakdown of the classical 2D rupture theory and that the linear approximation to $\tau(v)$ in Eq.~\eqref{eq:effective_linearization} opens the way to the development of the UST. The latter theory has been comprehensively shown to be in quantitative agreement with extensive numerical data up to rupture propagation speeds close to $c_{_{\rm R}}$. Yet, very close to $c_{_{\rm R}}$, the UST breaks down in itself, most likely due to nonlinearity in $\tau(v)$. Consequently, frictional rupture crosses $c_{_{\rm R}}$ and propagates inside the ``forbidden'' super-Rayleigh range.

\begin{figure}[ht!]
\centering
\includegraphics[width=\columnwidth]{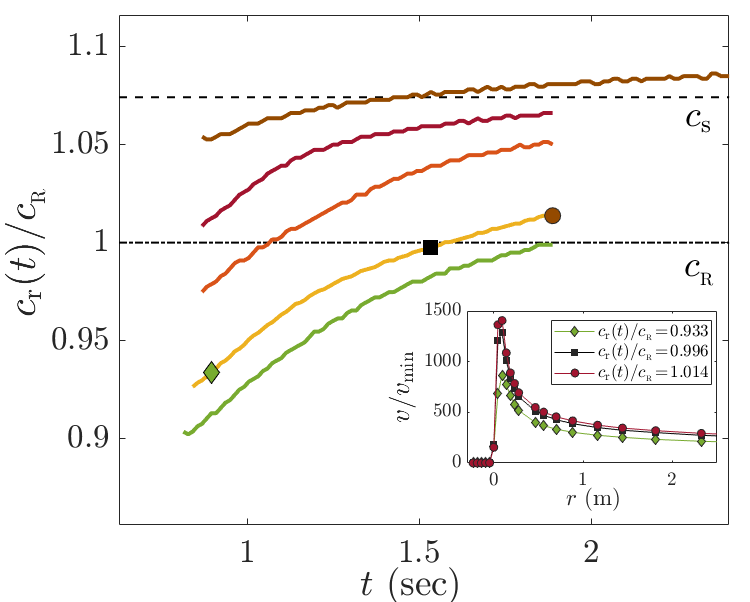}
\caption{{\bf Super-Rayleigh and super-shear frictional rupture propagation}. $c_{\rm r}(t)$ obtained for different prestress levels $\tau_{\rm d}$ (increasing from the bottom curve to the top one), see {\color{blue}{\em SI Appendix}}, revealing smooth and continuous frictional rupture propagation into and inside the super-Rayleigh range. The top curve reveals a continuous super-shear crossover (note that $c_{\rm s}\!=\!1.074c_{_{\rm R}}$ for the used elastic moduli). (inset) $v(r)$ for the three $c_{\rm r}$ values marked on the second curve from the bottom, see legend. The smallest $c_{\rm r}$ is somewhat below $c_{_{\rm R}}$ (the same as in Fig.~\ref{fig:unconventional_singularity_high}a), the middle one is essentially at $c_{_{\rm R}}$ and the largest features $c_{\rm r}\!>\!c_{_{\rm R}}$. Note that there is no divergence and change in the sign of $v(r)$ upon crossing $c_{_{\rm R}}$, in contrast to the classical theory prediction in Eq.~\eqref{eq:vr}, once $\xi\!=\!-1/2$ is substituted therein.}
\label{fig:super_Rayleigh}
\end{figure}

An important corollary of this major observation is that there should be no discontinuous/sharp super-shear transition in our numerical simulations, i.e., that rupture also smoothly and continuously propagates into the super-shear regime. This expectation is explicitly supported by the $c_{\rm r}(t)$ curve for the largest $\tau_{\rm d}$ value in Fig.~\ref{fig:super_Rayleigh}, yet again demonstrating the potentially profound implications of frictional rate dependence for fast earthquake propagation. Continuous rupture propagation through the super-Rayleigh range into the super-shear one has been documented in experiments~\cite{Kammer2018} and numerical simulations~\cite{festa2006influence,lu2009analysis,liu2014progression,bizzarri2016near,liang2022paucity}, where it has been termed the ``direct transition'', though it has not been fully understood theoretically.

\vspace{-0.2cm}
\section{S\lowercase{ummary and conclusion}}

The overarching theme of this work is to revisit the assumptions underlying the classical 2D theory of frictional rupture, with a focus on the roles played by a nonlinear rate-dependence frictional strength $\tau(v)$, which is experimentally established. While we consider a broad range of rupture propagation speeds, special attention is given to high propagation speeds, where the classical theory predicts the existence of a ``forbidden'' super-Rayleigh range, which raises the fundamental question of how rupture can transition into the super-shear range. We explain that the classical theory is based on the leading-order approximation of $\tau(v)$, simply a rate-independent constant (the residual stress $\tau_0$), combined with the assumption that the linear elastodynamic bulk equations admit a power-law series solution, with a singular contribution. The latter is the famous $\xi\=-1/2$ singularity of the classical theory.

By relaxing the first assumption, i.e., by considering a perturbative expansion of $\tau(v)$ up to the linear term in $v$, we develop the UST, which offers analytic predictions for frictional rupture with a non-universal singularity order $\xi\!\ne\!-1/2$. Using extensive numerical simulations, we demonstrated that the UST is in quantitative agreement with numerical observations over a broad range of propagation speed with $\xi\!\ne\!-1/2$, including close to the Rayleigh wave-speed $c_{_{\rm R}}$. These results demonstrate the breakdown the classical 2D theory of frictional rupture due to the rate-dependence of frictional strength $\tau(v)$, approximated to linear order in the slip rate $v$. Recent experiments also support the UST in the sub-Rayleigh range~\cite{paglialunga2022scale,paglialunga2024frictional,fryer2024effect}.

Yet, as $c_{_{\rm R}}$ is approached, the range of validity of the linear approximation to $\tau(v)$ shrinks and the UST breaks down in itself due to nonlinearity in $\tau(v)$, which most likely invalidates the power-law series solution assumption, i.e., a singular exponent $\xi$ can no longer be selected. As a consequence, frictional rupture smoothly and continuously propagates into and inside the so-called ``forbidden'' super-Rayleigh range. This finding has an immediate and important implication: in contrast to prediction of the classical theory, frictional rupture can also smoothly and continuously propagate into the super-shear range, i.e., surpassing the shear wave-speed $c_{\rm s}$ with no discontinuous transition, as we explicitly demonstrated. A similar behavior has been documented in a laboratory experiment~\cite{Kammer2018}.

Our results about the nature of the emergence of super-shear frictional rupture are not necessarily universal, i.e., some aspects of them might depend on properties of $\tau(v,\ldots)$, where the ellipsis stands for interfacial state fields. Future work should elucidate what properties of the interfacial constitutive law give rise to qualitative deviations from the predictions of the classical 2D rupture theory. On the other hand, our results show that the classical 2D rupture theory --- and specifically the existence of a ``forbidden'' super-Rayleigh range --- are not universal. Future work should consider applications of the developed theory and its predictions to geophysical field and laboratory observations, e.g., in relation to recent observations~\cite{Kammer2018,chounet2018global,bao2019early}, as well as extensions to 3D.\\

\vspace{1cm}

{\em Author contributions} --- E.B.~conceived and designed research, developed the theory and wrote the paper; A.P.~performed all numerical simulations, the comparison to the theory and generated all figures; F.B.~contributed to the theory development and provided support to the simulations; all authors discussed the results and contributed to the writing.\\

{\em Acknowledgements} --- We thank Efim Brener and Alina Shafir for useful discussions. E.B.~acknowledges support from the Minerva Foundation (with funding from the Federal German Ministry for Education and Research) and the Harold Perlman Family. F.B.~acknowledges support from the Research Council of Norway through its Researcher Project for Young Talents program (UNLOC, project 345008). The computations were carried out at Weizmann Institute of Science, using the Faculty of Chemistry's high-performance computing facility CHEMFARM, which is supported in part by the Ben May Center for Chemical Theory and Computation.

\clearpage

\onecolumngrid
\begin{center}
	\textbf{\large {\color{blue}{\em Supplemental Information (SI) Appendix}} for: ``Breakdown of the classical rupture theory and earthquake propagation in the "forbidden" super-Rayleigh range''}
\end{center}

\setcounter{equation}{0}
\setcounter{figure}{0}
\setcounter{section}{0}
\setcounter{table}{0}
\setcounter{page}{1}
\makeatletter
\renewcommand{\theequation}{S\arabic{equation}}
\renewcommand{\thefigure}{S\arabic{figure}}
\renewcommand{\thesection}{S-\arabic{section}}
\renewcommand{\thetable}{S-\arabic{table}}
\renewcommand*{\thepage}{S\arabic{page}}

\twocolumngrid

The goal of this document is to provide technical details about the theoretical and numerical results presented in the manuscript.

\section{T\lowercase{he unconventional singularities theory (\uppercase{UST})}}
\label{sec:theory_SM}

The development of the theory starts, as explained in the manuscript, from the linear momentum balance equation under plane-strain conditions
\begin{equation}
\rho\,\ddot{\mat{u}} - \nabla\cdot\mat{\sigma} = 0 \ ,
\label{momentum}
\end{equation}
valid for each of the bodies forming the fault. Here, $\rho$ being the mass density, $\mat{\sigma}$ the stress tensor, $\mat{u}(x,y,t)$ is the 2D displacement vector and each superposed dot denotes a partial time derivative. Based on the Helmholtz decomposition theorem, any vector field can be expressed as the sum of two contributions derived from a scalar potential $\phi$ and a vectorial potential $\mat{\Psi}$, respectively, according to
\begin{equation}
\mat{u} = \nabla\phi + \nabla \times \mat{\Psi} \ .
\label{helmholtz}
\end{equation}
Using Hooke's law for isotropic elastic bodies, the Helmholtz decomposition in Eq.~(\ref{helmholtz}) can be used to rewrite the momentum balance equation as a set of two wave equations
\begin{equation}
c_{\rm d}^2\nabla^2\phi = \ddot{\phi} \ , \qquad\qquad  c_{\rm s}^2\nabla^2\mat{\Psi} = \ddot{\mat{\Psi}} \ .
\label{waves}
\end{equation}

Equations (\ref{waves}) imply that isovolumetric deformation propagates at the shear wave-speed $c_{\rm s}$, whereas irrotational deformation travels at the dilatational wave-speed $c_{\rm d}$, both expressed in terms of Poisson's ratio $\nu$ and the shear modulus $\mu$ (appearing in Hooke's law). Next, we consider a dynamic earthquake rupture propagating in the $x$ direction at speed $c_{\rm r}(t)$ along the fault separating the two linear elastic solids (located at $y\=0$).

We introduce a Cartesian frame of reference defined by the unit vectors $\hat{\bm x}$ and $\hat{\bm y}$, whose directions are respectively parallel to the rupture propagation direction and normal to the fault plane, and assume plane-strain conditions (as noted above)) such that $\mat{\Psi}\=\psi\hat{\bm z}$, with the unit vector $\hat{\bm z}$ being perpendicular to the $\hat{\bm x}\!-\!\hat{\bm y}$ plane. We invoke the Lorentz transform, which preserves the shape of wave equations and reads
\begin{align}
&\frac{x-c_{\rm r}t}{\alpha_{\rm s,d}} = r_{\rm s,d}\cos(\theta_{\rm s,d})\ , &y = r_{\rm s,d}\sin(\theta_{\rm s,d}) \ ,
\label{equ:relativistic}
\end{align}
for a Cartesian frame of reference co-moving with the rupture edge. In Eq.~\eqref{equ:relativistic}, we used the definition $\alpha_{\rm s,d}\=\sqrt{1-c^2_{\rm r}/c^2_{\rm s,d}}$. Equations~(\ref{equ:relativistic}) actually correspond to two ``relativistic'' frames of reference, where distances along the propagation direction significantly contract as the rupture speed approaches $c_{\rm s,d}$, respectively. One can similarly define two co-moving relativistic polar system of coordinates
\begin{align}
\begin{split}
r_{\rm s,d} &=\frac{r}{\alpha_{\rm s,d}}\sqrt{1-c_{\rm r}^2\sin^2(\theta)/c^2_{\rm s,d}} \equiv \frac{r}{\alpha_{\rm s,d}}\gamma_{\rm s,d} \ ,\\ \;\; \theta_{\rm s,d} &= \tan^{-1}\Big(\alpha_{\rm s,d}\tan(\theta)\Big) \ ,
\label{lorentz}
\end{split}
\end{align}
with $r$ denoting the distance from the rupture edge and $\theta$ the angle relative to the propagation direction. Next, we assume that time variations are slaved to the smoothly time-varying rupture speed $c_{\rm r}(t)$, such that steady-state solutions can be searched in the co-moving frame of reference, i.e. $\partial_t\=0$. In fact, as shown in~\cite{Freund1998} and noted in the manuscript, the singular contribution to the solution (which is our main focus, see below and the manuscript), is valid for any $c_{\rm r}(t)$, without invoking the quasi-steady-state assumption.

Consequently, the wave equations (\ref{waves}) become a set of Laplace equations for $\phi$ and $\psi$ (recall that $\mat{\Psi}\=\psi\hat{\bm z}$). As explained in the manuscript, it is then {\bf assumed} that these equations admit a power-law series solution of the form $\mat{\nabla u}\= r_{\rm d,s}^{\xi} \mat{\mathcal{F}}(\theta_{\rm d,s},\xi)$, with $\mat{\mathcal{F}}(\cdot)$ being a function to be determined, leading to a solution of the form
\begin{align}
\begin{split}
\phi(r_{\rm d},\theta_{\rm d}) &= r_{\rm d}^{\xi+2}\Big(A\sin\left[(\xi+2)\theta_{\rm d}\right]+B\cos\left[(\xi+2)\theta_{\rm d}\right]\Big) \ ,\\
\psi(r_{\rm s},\theta_{\rm s}) &= r_{\rm s}^{\xi+2}\Big(C\sin\left[(\xi+2)\theta_{\rm s}\right]+D\cos\left[(\xi+2)\theta_{\rm s}\right]\Big) \ .
\label{general_solution}
\end{split}
\end{align}
The relations between the constants $A, B, C, D$ and the exponent $\xi$ can be determined by the boundary conditions. Hereafter, we assume mode-II symmetry, i.e., $u_x(r_{\rm s,d},-\theta_{\rm s,d})\= -u_x(r_{\rm s,d},\theta_{\rm s,d})$, which implies that $B\=C\=0$. The resulting mode-II displacement solution reads
\begin{align}
\begin{split}
u_x\!=\!(\xi+2) \Big(&r_{\rm d}^{\xi+1}\alpha^{-1}_{\rm d} A\sin\left[(\xi+1)\theta_{\rm d}\right] \\ - &r_{\rm s}^{\xi+1} D\sin\left[(\xi+1)\theta_{\rm s}\right]\Big) \ , \\
u_y\!=\!(\xi+2) \Big(&r_{\rm d}^{\xi+1} A\cos\left[(\xi+1)\theta_{\rm d}\right]\\ - &r_{\rm s}^{\xi+1}\alpha^{-1}_{\rm s} D\cos\left[(\xi+1)\theta_{\rm s}\right]\Big) \ .
\label{displacement_solution}
\end{split}
\end{align}

Consider then a constant normal traction along the fault and assume that the shear traction is dominated by a singular stress contribution of the form
\begin{equation}
    \sigma_{xy}(r,\theta_{\rm s,d}=0)=K_{\rm II}^{(\xi)}r^{\xi}/\sqrt{2\pi} \ ,
\end{equation}
where $K_{\rm II}^{(\xi)}$ is a generalized \textit{stress intensity factor}. Using the above, the ratio between $A$ and $D$ can be determined and one ends up with
\begin{align}
\begin{split}
&u_x(r,\theta) = r^{\xi+1}\frac{K_{\rm II}^{(\xi)}}{(\xi+1)\,\mu\,R(c_{\rm r})\sqrt{2\pi}} \,\times\\ &\Big(\!2\gamma_{\rm d}^{\xi+1}\alpha_{\rm s} \sin\!\left[(\xi+1)\theta_{\rm d}\right] - \gamma_{\rm s}^{\xi+1} \alpha_{\rm s}(1+\alpha^{2}_{\rm s})\sin\!\left[(\xi+1)\theta_{\rm s} \right]\!\Big) \ , \\
&u_y(r,\theta) = r^{\xi+1}\frac{K_{\rm II}^{(\xi)}}{(\xi+1)\,\mu\,R(c_{\rm r})\sqrt{2\pi}} \,\times \\ &\Big(\!2\gamma_{\rm d}^{\xi+1}\alpha_{\rm d}\alpha_{\rm s} \cos\!\left[(\xi+1)\theta_{\rm d}\right] - \gamma_{\rm s}^{\xi+1}(1+\alpha^{2}_{\rm s})\cos\!\left[(\xi+1)\theta_{\rm s}\right]\!\Big) \ ,
\label{SIF_displacement_solution}
\end{split}
\end{align}
for the displacement, with
\begin{align}
\begin{split}
\dot{u}_x(r,\theta) &= \frac{K^{(\xi)}_{\rm II}}{\sqrt{2\pi}} r^{\xi}\frac{c_{\rm r}\,\alpha_{\rm s}}{\mu\,R(c_{\rm r})}\,\times \\ &\left[-2\gamma_{\rm d}^{\xi} \sin(\xi\theta_{\rm d})+(1+\alpha^{2}_{\rm s})\gamma_{\rm s}^{\xi}\sin(\xi\theta_{\rm s})\right] \ ,\\
\dot{u}_y(r,\theta) &= \frac{K^{(\xi)}_{\rm II}}{\sqrt{2\pi}} r^{\xi}\frac{c_{\rm r}}{\mu\,R(c_{\rm r})} \,\times\\ &\left[-2\alpha_{\rm d}\alpha_{\rm s}\gamma_{\rm d}^{\xi}\cos(\xi\theta_{\rm d})+(1+\alpha^2_{\rm s})\gamma_{\rm s}^{\xi}\sin(\xi\theta_{\rm s})\right] \ ,
\label{SIF_velocity_solution}
\end{split}
\end{align}
for the particle velocity, and
\begin{align}
\begin{split}
\sigma_{xy}(r,\theta) &= \frac{K^{(\xi)}_{\rm II}}{\sqrt{2\pi}} r^{\xi}\frac{1}{R(c_{\rm r})} \,\times\\&\left[ 4\alpha_{\rm d}\alpha_{\rm s} \gamma_{\rm d}^{\xi} \cos(\xi\theta_{\rm d})- (1+\alpha^{2}_{\rm s})^2\gamma_{\rm s}^\xi\cos(\xi\theta_{\rm s})\right] \ ,\\
\sigma_{yy}(r,\theta) &= \frac{K^{(\xi)}_{\rm II}}{\sqrt{2\pi}} r^{\xi} \frac{2\alpha_{\rm s}(1+\alpha_{\rm s}^2)}{R(c_{\rm r})} \,\times\\&\left[-\gamma_{\rm d}^{\xi} \sin(\xi\theta_{\rm d})+\gamma_{\rm s}^{\xi}\sin(\xi\theta_{\rm s})\right] \ ,
\label{SIF_stress_solution}
\end{split}
\end{align}
for the stress. The Rayleigh function reads
\begin{equation}
R(c_{\rm r})=4\alpha_{\rm d}\alpha_{\rm s} - (1+\alpha^2_{\rm s})^2 \ ,
\label{eq:Rayleigh}
\end{equation}
featuring a non-trivial root $R(c_{\rm r}\=c_{_{\rm R}})\=0$ that defines the Rayleigh wave-speed $c_{_{\rm R}}$. Equations~(1)-(3) in the manuscript then follow, where the slip rate is $v(r,t)\=2\dot{u}_x(r,\theta\=\pi,t)$.

\begin{figure}
\includegraphics[width=0.3\textwidth]{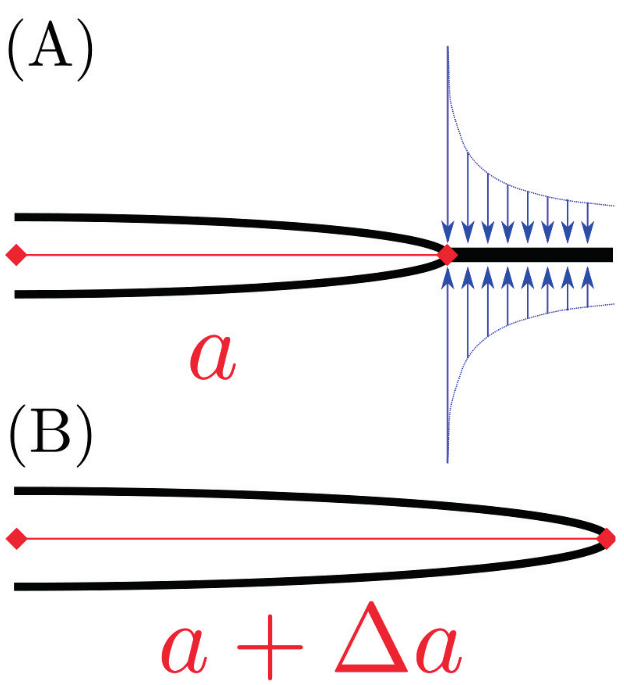}
\vspace{-0.25cm}
\caption{Illustration of Irwin's original approach for mode-I (tensile) rupture~\cite{Irwin1957} to compute the change of potential energy due to tensile crack advance, i.e., between states (A) and (B). In order to keep the boundary surfaces unchanged, Irwin assumed that a pair of opposite normal tractions (blue arrows) maintains the crack closed over the surface increment $\Delta a$ and mimics the effect of the material's cohesion. We follow the same approach to integrate the energy release rate of mode-II rupture governed by unconventional singularities.}
\label{fig:irwin}
\end{figure}

We next consider the energy associated with rupture propagation. In the {\bf co-moving frame of reference}, the mechanical energy corresponds to the potential energy $\Pi$. This useful property enables one to apply the original approach proposed by Irwin~\cite{Irwin1957} for quasi-static crack, see Fig.~\ref{fig:irwin}, to our problem. Irwin considered a pair of opposite tractions applied over a distance $\Delta{a}$ to initially maintain the crack surfaces closed. In this context, the change in potential energy $\mathrm{d}\Pi(\Delta{a})$ resulting from rupture propagation over $\Delta{a}$ can be expressed as the work done by surface tractions $\mat{t}\=\mat{\sigma} \cdot \mat{n}$ along the fault
\begin{equation}
    \mathrm{d}\Pi(\Delta a) = -\frac{1}{2}\int_{\Delta a}\mathrm{d}(\mat{t}\cdot\mat{u})\,\mathrm{d}\mathcal{S} \ .
    \label{equ:epot_irwin}
\end{equation}
Note that Eq.~(\ref{equ:epot_irwin}) integrates only the flux of mechanical energy towards the near-edge portion of the fault plane of size $\Delta{a}$ and not over the entire fault surface $\mathcal{S}$.

Following Irwin's approach, we consider a pair of opposite shear tractions that exactly balance the singular shear stress existing ahead of the rupture edge, i.e., $t_x\=\sigma_{xy}(r_{\rm s,d},\theta_{\rm s,d}=0)$, and prevent frictional slip. We underline that Eq.~(\ref{equ:epot_irwin}) has to be evaluated independently for the isovolumetric (shear) and irrotational (dilatational) parts of the solution as they reside in different frame of reference. As a result, the integration of Eq.~(\ref{equ:epot_irwin}) should also account for the work of tensile traction arising behind the rupture edge, since $t_y\=\sigma_{yy}(r_{\rm s,d},\theta_{\rm s,d}\=\pi)$ has finite isovolumetric and irrotational contributions and only their sum is zero. Keeping these properties in mind, the mechanical energy released by growing the rupture over $\Delta{a}$ can be obtained as
\begin{align}
\begin{split}
    &\mathrm{d}\Pi(\Delta a) = \\ &\sum_{i={\rm s,d}}-\int_0^{\Delta{a}}\!\!\left[ \sigma_{xy}\Big(r_{i}\!=\!\mathcal{X};\theta_{i}\!=\!0\Big)\;u_{x}\Big(r_i\!=\!\Delta a -\mathcal{X};\theta_i\!=\!\pi\Big)\right. \\ & \left. - u_y\Big(r_i\!=\!\mathcal{X};\theta_i\!=\!0\Big)\;\sigma_{yy}\Big(r_i\!=\!\Delta{a} -\mathcal{X};\theta_i\!=\!\pi\Big) \right] \mathrm{d}\mathcal{X}\\
    &=-\frac{\sin(-\pi\xi)\,c^2_{\rm r}}{2\sqrt{\pi}(\xi+1)\,\mu\,R^2\,c^2_{\rm s}}\frac{(\Delta{a})^{2(\xi+1)}\;\Gamma(\xi+1)}{4^{\xi+1}\;\Gamma(\xi+\frac{3}{2})} \,\times\\ &\Big(4\alpha^2_{\rm s}\alpha^{2(\xi+1)}_{\rm d}-\alpha_{\rm s}^{2(\xi+1)}(1+\alpha^2_{\rm s})^2\Big)\left[K^{(\rm \xi)}_{\rm II}\right]^2 \ ,
\label{equ:delta_epot}
\end{split}
\end{align}
where $\Gamma$ denotes the Gamma function.

\subsection{The classical rupture theory and generalized energy balance}

Before we derive the generalized energy balance for frictional rupture characterized by an unconventional singularity, let us briefly obtain the results corresponding to the classical rupture theory~\cite{Freund1998}. As discussed extensively in the manuscript, the classical rupture theory is recovered when Eq.~(6) in the manuscript is truncated to leading order (the rate-independent limit of the frictional strength $\tau$), i.e., $\tau(v)\!\simeq\!\tau_0$, where $\tau_0$ is a constant residual stress. Imposing the boundary condition $\sigma_{xy}(r,\theta\=\pm\pi,t)\=\tau_0$, we conclude that the singular contribution in Eqs.~(\ref{SIF_stress_solution}) must vanish behind the edge, which is ensured by the  conventional singularity, $\xi\=-1/2$. This is equivalent to setting the RHS of Eq.~(7) in the manuscript to zero.

Using then $\xi\=-1/2$ in Eqs.~(\ref{SIF_stress_solution}), one recovers the classical square-root singular fields~\cite{Freund1998}
\begin{align}
\begin{split}
\sigma_{xy}(r,\theta) &= \frac{K_{\rm II}}{\sqrt{2\pi r}} \frac{1}{R} \left[ 4\alpha_{\rm d}\alpha_{\rm s} \frac{\cos\frac{1}{2}\theta_{\rm d}}{\sqrt{\gamma_{\rm d}}}- (1+\alpha^{2}_{\rm s})^2\frac{\cos\frac{1}{2}\theta_{\rm s}}{\gamma_{\rm s}}\right],\\
\sigma_{yy}(r,\theta) &= \frac{K_{\rm II}}{\sqrt{2\pi r}} \frac{2\alpha_{\rm s}(1+\alpha_{\rm s}^2)}{R}\left[\frac{\sin\frac{1}{2}\theta_{\rm d}}{\gamma_{\rm d}}-\frac{\sin\frac{1}{2}\theta_{\rm s}}{\gamma_{\rm s}}\right] \ ,
\label{LEFM_stress_solution}
\end{split}
\end{align}
where $K_{\rm II}\=K^{(\xi=-1/2)}_{\rm II}$. The computation of the energy release rate $G$~\cite{Freund1998} then follows from Eq.~(\ref{equ:delta_epot})
\begin{equation}
 G \equiv \lim_{\Delta{a} \to 0 }\left(\!-\frac{\mathrm{d}\Pi(\Delta{a})}{\Delta{a}}\right) = \frac{1-\nu^2}{E}A_{\rm II}(c_{\rm r},\xi\!=\!-1/2)\,K_{\rm II}^2 \ ,
 \label{equ:frac_err}
\end{equation}
where $E\=2\mu(1+\nu)$ is Young's modulus and~\cite{Freund1998}
\begin{equation}
 A_{\rm II}(c_{\rm r},\xi\!=\!-1/2) = \frac{c_{\rm r}^2\,\alpha_{\rm s}}{(1-\nu)R(c_{\rm r})\,c^2_{\rm s}} \ .
 \label{equ:conv_AII}
\end{equation}

To obtain the energy release rate $G$ in the generalized case of an unconventional singularity $\xi\!\ne\!-1/2$, note that $\Delta{a}$ can no longer be arbitrarily small and one needs to introduce a finite cohesive zone of size $\ell_{\rm c}$, as done in the manuscript. Consequently, we obtain
\begin{equation}
    G \!=\!\lim_{\Delta{a} \to \ell_{\rm c}}\left(\!-\frac{\mathrm{d}\Pi(\Delta a)}{\Delta{a}}\right) = \frac{1-\nu^2}{E}\Big[K^{(\rm \xi)}_{\rm II}\Big]^2A_{\rm II}(c_{\rm r},\xi)\,\ell_{\rm c}^{^{2\xi+1}} \ ,
\label{equ:unconv_err}
\end{equation}
which identifies with Eq.~(5) in the manuscript. The generalized function $A_{\rm II}(c_{\rm r},\xi)$ reads
\begin{align}
\label{equ:unconv_AII}
\begin{split}
    A_{\rm II}(c_{\rm r},\xi) &=  \left(\frac{c_{\rm r}}{c_s}\right)^2\left(4\alpha^2_s\alpha^{2(\xi+1)}_d-\alpha_s^{2(\xi+1)}(1+\alpha^2_s)^2\right)\\&\,\times\frac{\sin(-\pi\xi)}{(1-\nu)\left[R(c_{\rm r}) \right]^2}\frac{\Gamma(\xi+1)}{4^{\xi+1}\sqrt{\pi}(\xi+1)\;\Gamma(\xi+3/2)} \ .
\end{split}
\end{align}
Note that the Rayleigh function $R(c_{\rm r})$ enters the denominator of Eq.~(\ref{equ:unconv_AII}) quadratically, in contrast to the classical result in Eq.~\eqref{equ:conv_AII}, where it appears linearly in the denominator. As discussed in the manuscript, the latter --- specifically $R(c_{\rm r}\=c_{_{\rm R}})\=0$ and the sign change of $R(c_{\rm r})$ for $c_{\rm r}\!>\!c_{_{\rm R}}$ --- is the origin of the ``forbidden'' super-Rayleigh range of propagation wave-speeds in the classical theory. On the other hand, the generalized $A_{\rm II}(c_{\rm r},\xi)$ in Eq.~\eqref{equ:unconv_AII} does not change sign across $c_{_{\rm R}}$ due to $R(c_{\rm r})$ for $\xi\!\ne\!-1/2$.

Finally, we consider the so-called \textit{breakdown energy} that is defined in the manuscript as
\begin{equation}
\label{eq:Gf}
G_{\rm f}(\delta)\equiv \int_0^{\delta}[\tau(\delta')-\tau(\delta)]\,d\delta' \ ,
\end{equation}
which integrates the local frictional dissipation on top of residual friction, where $\delta(r)\= 2u_x(r,\theta\=\pi)$ is the fault slip. The edge-localized dissipation $G_{\rm c}$ corresponds to $G_{\rm f}(\delta\=\delta_{\rm c})\=G_{\rm c}$, where $\delta_{\rm c}$ is the fault slip associated with $\ell_{\rm c}$. Using then Eqs.~(\ref{SIF_displacement_solution}) and (\ref{SIF_stress_solution}), one obtains $\delta(r)\!\sim\!r^{\xi+1}$ and $\tau(r)\=\sigma_{xy}(r,\theta\=\pi)\!\sim\!r^{\xi}\!\sim\! \delta^{\frac{\xi}{\xi+1}}$ behind the rupture edge, such that the breakdown energy follows
\begin{equation}
\label{eq:GF_scaling}
    G_{\rm f}(\delta) = G_{\rm c}\,\left(\delta/\delta_{\rm c}\right)^{\frac{1+2\xi}{1+\xi}} \ ,
\end{equation}
for $\delta\!>\!\delta_{\rm c}$, which identifies with Eq.~(4) in the manuscript. The exponent $\xi_{_{\rm G}}\!(\xi)\=(1+2\xi)/(1+\xi)$ is extensively discussed in the manuscript and below.

\section{N\lowercase{umerical simulations}}
\label{sec:numerics_SM}

\subsection{The boundary integral method approach}
\label{ssec:bi_method}

The linear elastic bodies forming the fault (located at $y\=0$) are assumed in our problem to be infinite in the direction perpendicular to it. Under this condition, the interfacial shear stress $\sigma_{xy}(x,t)$ can be computed using the boundary integral approach, where the fault boundary condition $\tau\=\sigma_{xy}(x,t)$ takes the form~\cite{das1980numerical}
\begin{equation}
\label{eq:BIM}
\tau[v(x,t),\phi(x,t)] = \tau_{\rm d} -\frac{\mu}{2c_{\rm s}}v(x,t) + s(x,t) \ .
\end{equation}
Here, $\tau_{\rm d}$ is the background stress (prestress) applied far from the fault (the subscript `d' stands for `driving') and $(\mu/2c_{\rm s})v(x,t)$ is the so-called radiation damping term, which depends on the local slip rate $v(x,t)$. $s(x,t)$ is a spatiotemporal convolutional integral that accounts for the long-range interaction of different parts of the fault, mediated by bulk deformation. In general, $s(x,t)$ does not admit an explicit real-space representation, but is rather expressed in Fourier space~\cite{Breitenfeld1998}.

The interfacial shear stress, corresponding to the RHS of Eq.~\eqref{eq:BIM}, is balanced at each point in time and fault position by the frictional strength $\tau[v(x,t),\phi(x,t)]$, which depends in addition to $v(x,t)$ on the internal state field $\phi(x,t)$ that admits its own evolution equation; both are specified in the next section. Once specified, we solve the problem  numerically using the open-source library {\em cRacklet}~\cite{roch2022cracklet}, which is based on a spectral representation of Eq.~\eqref{eq:BIM}. It involves a finite fault of size $W$ and periodic boundary conditions such that the basic fields in the problem satisfy $v(x,t)\=v(x+W,t)$ and $\phi(x,t)\=\phi(x+W,t)$. The problem formulation is completed once the initial conditions $v(x, t\=0)$ and $\phi(x, t\=0)$ are specified, to be discussed below.

\subsection{The fault constitutive relation\\ (the friction law)}
\label{ssec:friction_curve}

We employed a nonlinear rate-and-state dependent frictional strength $\tau[v(x,t),\phi(x,t)]$, which is based on extensive laboratory experiments~\cite{Marone1998a,Baumberger2006,Bar-Sinai2014}, taking the form~\cite{Brener2018}
\begin{eqnarray}
\label{eq:f}
\tau(v,\phi)/\sigma &=& \left[1+b\log\left(1+\phi/\phi^{\star}\right)\right] \times \\ \nonumber
&&\left[f_0/\sqrt{1+\left(v^{\star}\!/v\right)^2}+\alpha\,\log\left(1+{\abs v}/v^{\star}\right)\right]\ .
\end{eqnarray}
The values of all parameters appearing in Eq.~\eqref{eq:f} can be found in Table~\ref{tab:1}, along with the value of the employed normal stress $\sigma$ and of the linear elastic parameters. The internal state field $\phi(x,t)$, which quantifies the amount of real contact area, satisfies the differential equation (known as the `aging-law'~\cite{Marone1998a,Baumberger2006})
\begin{equation}
\partial_t\phi(x,t)\=1-\!\frac{v(x,t)\,\phi(x,t)}{D}\, ,
\label{eq:phi}
\end{equation}
where $D$ is the characteristic slip distance, whose value is also given in Table~\ref{tab:1}.

\begin{figure}
\includegraphics[width=0.42\textwidth]{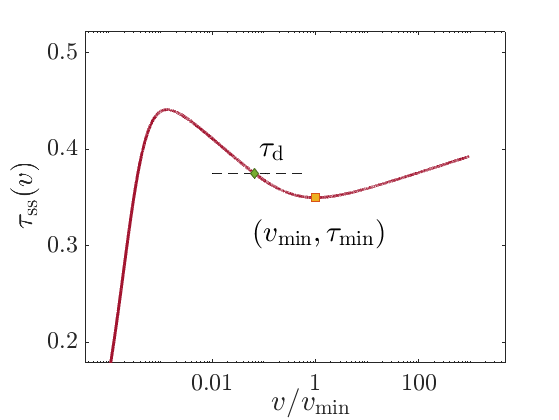}
\vspace{-0.25cm}
\caption{The steady frictional resistance $\tau_{\rm ss}(v)$ in Eq.~\eqref{eq:fss}. The orange square marks the position of the friction curve minimum. The horizontal dashed line denotes the background stress. The green diamond corresponds to the solution $\tau(v_{\rm vw},\phi_{\rm ss})\!=\!\tau_{\rm d}$ on the velocity-weakening (vw) branch of the steady-state friction curve.}
\label{fig:f1sm}
\end{figure}

Under persistent sliding at a slip rate $v_{\rm ss}$, Eq.~\eqref{eq:phi} attains a steady state of the form $\phi\=D/v_{\rm ss}$. When the latter is substituted in $\tau(v,\phi)$ of Eq.~\eqref{eq:f}, the steady-state friction curve is obtained
\begin{eqnarray}
\label{eq:fss}
\tau_{\rm ss}(v)/\sigma &\!=\!& \left[1+b\log\left(1+D/(v_{\rm ss}~ \phi^{\star})\right)\right] \times \\ \nonumber
&&\left[f_0/\sqrt{1+\left(v^{\star}\!/v_{\rm ss}\right)^2}+\alpha\,\log\left(1+{\abs {v_{\rm ss}}}/v^{\star}\right)\right]\ ,
\end{eqnarray}
which is $N$ shaped, as highlighted in the manuscript. It is plotted in Fig.~\ref{fig:f1sm}, where the local minimum at $(v_{\rm min}, \tau_{\rm min})$ is marked. The characteristic slip rate $v_{\rm min}\=3.4624\times 10^{-5}$ m/s and characteristic frictional strength $\tau_{\rm min}\=0.3498$ MPa are used to nondimensionalize various quantities hereafter and in the manuscript.

We employed background stresses in the range $\tau_{\rm d}/\tau_{\rm min}\!\in\![1.02, 1.078]$. The results plotted in Fig.~1 in the manuscript correspond to $\tau_{\rm d}/\tau_{\rm min}\=1.04$ and those in Fig.~2 therein to $\tau_{\rm d}/\tau_{\rm min}\=1.071$. The curve in Fig.~4 in the manuscript correspond to $\tau_{\rm d}/\tau_{\rm min}\=1.070, 1.071, 1.073, 1.075, 1.078$, from bottom to top.
\begin{table}
\centering
\begin{tabular}{|c|c|c|}
\hline
$\sigma$ & $1$  & MPa\\
	\hline
$\nu$ & $0.33$  &--\\
	\hline
$\mu$            & $9$&	GPa\\
\hline
$c_{\rm s} $      &$2739$  &m/s\\
\hline
 $D$              &$5\times 10^{-7}$ & m\\
 \hline
  $v^*$           &$10^{-8}$  &m/s\\
  \hline
$\phi^*$         &$0.05$ & s\\
\hline
$\alpha$         & $0.0075$	&--\\
\hline
$b$              &$0.1$  &--\\
\hline
  $f_0$          &$0.28$ & --\\
	\hline
\end{tabular}
\caption{The normal stress $\sigma$, linear elastic bulk parameters, and frictional parameters, cf.~Eqs.~\eqref{eq:f}-\eqref{eq:phi}, used in the computer simulations.}
\label{tab:1}
\end{table}

\subsection{Rupture nucleation procedure}
\label{ssec:nucleation}

\begin{figure*}[htp]
\includegraphics[scale=0.45]{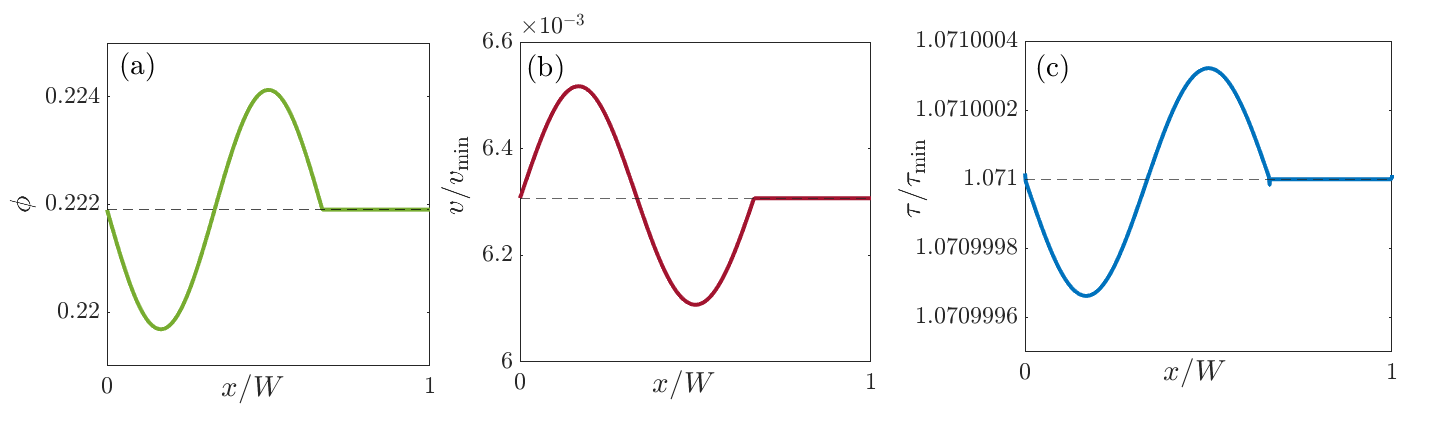}
\vspace{-0.25cm}
    \caption{(a) The initial perturbed state field $\phi(x,t\!=\!0^+)$, see Eq.~\eqref{eq:1}, (b) the corresponding slip rate $v(x,t\!=\!0^+)$ and (c) the corresponding frictional strength $\tau(x,t\!=\!0^+)$, for $\tau_d/\tau_{\rm min}\!=\!1.071$. The dashed horizontal lines  correspond to  $D/v_{\rm vw}, v_{\rm vw}$ and $\tau_{\rm d}$, respectively.}
\label{fig:f2sm} 	
\end{figure*}
As explained above, Eqs.~\eqref{eq:BIM}-\eqref{eq:phi} for $v(x,t)$ and $\phi(x,t)$ require initial conditions. The latter, which are also used to nucleate rupture, are formulated in two steps. First, for a given $\tau_{\rm d}$, we set $v(x,t\=0)\=v_{\rm vw}$ and $\phi(x,t\=0)\=\phi_{\rm ss}\=D/v_{\rm vw}$ to be space independent, where $v_{\rm vw}$ is the solution of $\tau(v_{\rm vw},\phi_{\rm ss})\=\tau_{\rm d}$ on the velocity-weakening (vw) branch of the steady-state friction curve (illustrated in Fig.~\ref{fig:f1sm} by a green diamond). Second, at $t\=0^+$, we introduce a perturbation to $\phi$ of the form
\begin{equation}
 \phi(x,t=0^+) =\left[1+\epsilon\,\sin(2\pi x/w)\right]D/v_{\rm vw} \,,
\label{eq:1}
\end{equation}
for $0\!\le\!x\!\le\!w$ and $\phi(x,t=0^+)\=D/v_{\rm vw}$ otherwise. Here, $\epsilon\=-0.01$ and $w$ is the perturbation width, and $v(x,t\=0^+)$ is set as if it is slaved to $\phi(x,t\=0^+)$, see Fig.~\ref{fig:f2sm}.

The above-described nucleation procedure gives rise to a pair of counter-propagating ruptures. We focus throughout this work on the {\bf right-propagating rupture}. Yet, due to the employed periodic boundary conditions in the fault direction, the left-propagating rupture eventually interacts with the right-propagating one. Since our entire analysis is performed for an isolated rupture, we increased the propagation distance and time of the right-propagating rupture {\bf prior} to any interaction with the left-propagating one through the periodic boundary conditions by choosing $w\=2W/3$ and a large enough $W$ (specifically, set to $W\=23136$ m). We verified that indeed our analysis of the right-propagating rupture is not affected by any interaction with the left-propagating rupture.

\vspace{-0.45cm}
\section{C\lowercase{omparing the theoretical predictions and the numerical observations}}

In this section, we provide additional details on the comparison between the theoretical predictions and the numerical observations for the right-propagating ruptures, giving rise to the results presented in Figs.~1-3 in the manuscript. In so doing, we also report the values of some of the extracted parameters not reported in the manuscript.

To test Eq.~\eqref{eq:GF_scaling} for the breakdown energy $G_{\rm f}(\delta)$ (Eq.~(4) in the manuscript), we consider the time dependence of the fault slip $\delta(x_i,t)$ and stress $\tau(x_i,t)$ at a fixed fault location $x_i$. Since the fault features a small, yet non-vanishing, background slip rate $v_{\rm vw}$, we actually use $G_{\rm f}(\delta)\= \int_{\delta_i}^{\delta}[\tau(\delta')\!-\!\tau(\delta)]\,d\delta'$, where $\delta_i$ corresponds to the slip accumulated prior to the arrival of the rupture to $x_i$.
\begin{figure}[htp]
\includegraphics[scale=0.42]{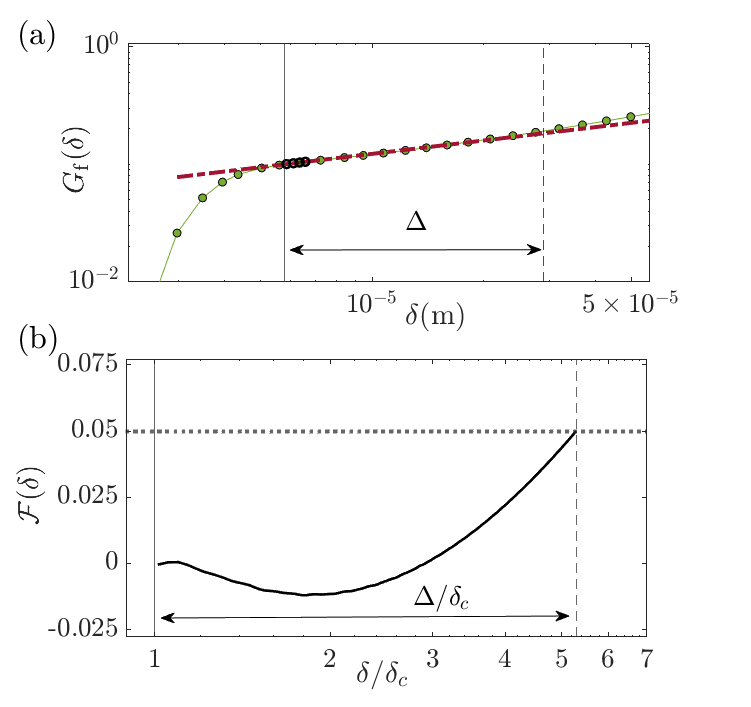}
\vspace{-0.25cm}
\caption{Estimating $\delta_{\rm c}$ and $\Delta$ (the data presented here correspond to Fig.~1 in the manuscript). (a) $G_{\rm f}(\delta)$ (green circles), as defined in the text, in a double-logarithmic representation. The data are logarithmically spaced for clarity. The $4$ points over which the power-law fit parameters were calculated are denoted by thicker black circles. The dashed brown line illustrates the power-law behavior and the corresponding range, $\Delta$, is marked with a double arrow. (b) ${\cal F}(\delta)$ of Eq.~\eqref{eq:Gf_diff}, see text for details. The vertical solid line in both panels corresponds to $\delta_{\rm c}$, and the vertical dashed line to the right-most edge of the power-law range. The dotted horizontal line in panel (b) corresponds to the chosen threshold of $0.05$, cf.~Eq.~\eqref{eq:Gf_diff}.}	
\label{fig:f3sm}
\end{figure}

The characteristic slip $\delta_{\rm c}$ is estimated as the slip above which $G_{\rm f}(\delta)$ follows a power-law, which is significantly slower than the variation for $\delta\!<\!\delta_{\rm c}$ that corresponds to the strong weakening inside the cohesive zone, as illustrated in Fig.~\ref{fig:f3sm}a (and in Figs.~1b and~2b in the manuscript). With $\delta_{\rm c}$ at hand, we proceed to determine $\xi_{_{\rm G}}$ in Eq.~\eqref{eq:GF_scaling} (Eq.~(4) in the manuscript), as well as the range $\Delta$ of the power-law scaling. We fit $G_{\rm f}(\delta)$, in a double-logarithmic representation, over the $4$ data points nearest to $\delta_{\rm c}$ (marked by thicker black circles in Fig.~\ref{fig:f3sm}a) to extract $\xi_{_{\rm G}}$. Then, using $\xi_{_{\rm G}}$, $\delta_{\rm c}$ and $G_{\rm c}\=G_{\rm f}(\delta_{\rm c})$, we compute the range of $\delta$ values --- denoted by $\Delta$ --- for which the relative difference between the two sides of Eq.~\eqref{eq:GF_scaling} satisfies
\begin{equation}
\label{eq:Gf_diff}
  {\cal F}(\delta)=1-\frac{G_{\rm c}\left(\delta/\delta_{\rm c}\right)^{\xi_{_{\rm G}}}}{G_{\rm f}(\delta)}<0.05\,,\qquad \delta \in \Delta\, ,
 \end{equation}
see Fig.~\ref{fig:f3sm}b. The values of $G_{\rm c}$ for $\tau_{\rm d}/\tau_{\rm min}\=1.04$ are reported in Fig.~1h in the manuscript, and the values of $\Delta/\delta_{\rm c}$ and $\xi_{_{\rm G}}$ for all the studied background stresses are reported in Fig.~3a,b in the manuscript.

Next, we analyze the spatial variation of the slip rate. To that aim, we consider a snapshot $v(r,t_i)$, e.g., as plotted in Fig.~1d in the manuscript, where $r$ is the distance from the location of the stress peak and $t_i$ is a fixed time. We then fit the instantaneous snapshot to a power-law $v(r, t_i)\=a_v r^{\xi}$ using a double logarithmic representation, as in Fig.~1f therein, over the spatial range corresponding to $\Delta$. The extracted exponent $\xi$ is then compared to the value independently obtained from inverting $\xi_{_{\rm G}}(\xi)$ of Eq.~(4) in the manuscript, i.e., $\xi\=(1-\xi_{_{\rm G}})/(\xi_{_{\rm G}}-2)$. The comparison is discussed in the manuscript in relation to Fig.~1-3 therein. The stress intensity $K_{II}^{(\xi)}(t)$, extracted for various time points $t_i$ and plotted in Fig.~1g in the manuscript, is obtained by equating $a_v$ to the analytic solution in Eq.~(2) in the manuscript.
\begin{figure}[ht!]
\includegraphics[scale=0.49]{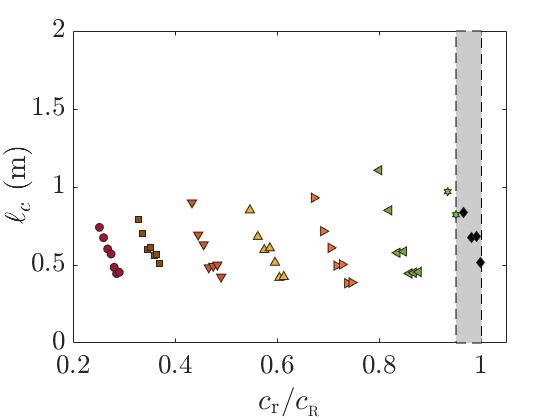}
\vspace{-0.25cm}
\caption{The cohesion zone size $\ell_{\rm c}$ vs.~$c_{\rm r}$ for various $\tau_{\rm d}$ values (see text for details). The symbols here are the same as in Fig.~3 in the manuscript and correspond to $\tau_{\rm d}/\tau_{\rm min}\!=\!1.04$ (leftmost, circles, cf.~Fig~1 in the manuscript), $1.045, 1.050, 1.055, 1.060, 1.065, 1.070$ and $1.071$ (rightmost, diamonds, cf.~Fig.~2).}
\label{fig:f4sm}	
\end{figure}

\begin{figure}[ht!]
\includegraphics[scale=0.49]{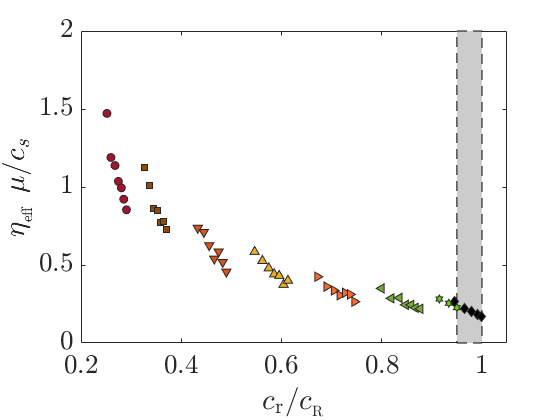}
\vspace{-0.25cm}
\caption{The effective viscosity $\eta_{\rm eff}$ vs.~$c_{\rm r}$ for various $\tau_{\rm d}$ values as in Fig.~\ref{fig:f4sm} (the same symbols are used).}	
\label{fig:f5sm}
\end{figure}

With $K_{II}^{(\xi)}$ and $G_{\rm c}$ at hand, we can test the rupture energy balance prediction in Eq.~(5) in the manuscript, with $A_{\rm II}(c_{\rm r},\xi)$ of Eq.~\eqref{equ:unconv_AII}. To this aim, the size $\ell_{\rm c}$ of the cohesion zone is computed by integrating the instantaneous rupture speed over the time it takes the fault slip to reach $\delta_{\rm c}$, i.e.,
\vspace{-0.25cm}
\begin{equation}
\label{eq:ellc}
\ell_{\rm c}=\int_{t_i}^{t_{\rm c}} c_{\rm r}(t') d t'\ ,
\vspace{-0.2cm}
\end{equation}
where $\delta(t_i)\=\delta_i$ and $\delta(t_{\rm c})\=\delta_{\rm c}$. The resulting $\ell_{\rm c}$ is plotted in Fig.~\ref{fig:f4sm} for all $\tau_{\rm d}$ simulations and times. The excellent agreement of the numerical data with the theoretical prediction in Eq.~(5) in the manuscript is demonstrated in Fig.~1h in the manuscript. Finally, for completeness, we plot in Fig.~\ref{fig:f5sm} the effective viscosity $\eta_{\rm eff}$ (the linear slope of $\tau(v)$ in the slip rate range corresponding to $\Delta$, as shown in Figs.~1e and 2a(inset) in the manuscript) for all simulations. These values were used in Fig.~3c in the manuscript to obtain $\xi^{(\eta)}$ using Eq.~(7) therein.

\vspace{-0.35cm}
\section{N\lowercase{umerical convergence}}
\vspace{-0.25cm}

In this section, we complement the previous two sections by demonstrating the robustness of our numerical results, i.e., the convergence of the simulations with respect to the spatial and temporal discretizations, $\Delta{x}$ and $\Delta{t}$, respectively. The latter two are related via the Courant-Friedrichs-Lewy (CFL) constant $\beta\=c_{\rm s}\Delta{t}/\Delta{x}\!<\!1$. In our simulations, $\beta\=0.2$, and $\Delta{t}$ follows from the spatial resolution. Given the very large fault size $W\=23136$ m, already stated above, fully resolving the fields inside the cohesion zone is challenging; yet, we confidently determine its edge, which allows to extract $\ell_{\rm c}$, as discussed above. Since we test the broad range of theoretical predictions for $\delta\!>\!\delta_{\rm c}$ (correspondingly, $r\!>\!\ell_{\rm c}$), i.e., out of the cohesive zone, this is enough for our purposes.
\begin{figure}[htbp]
\centering
\includegraphics[width=0.5\textwidth]{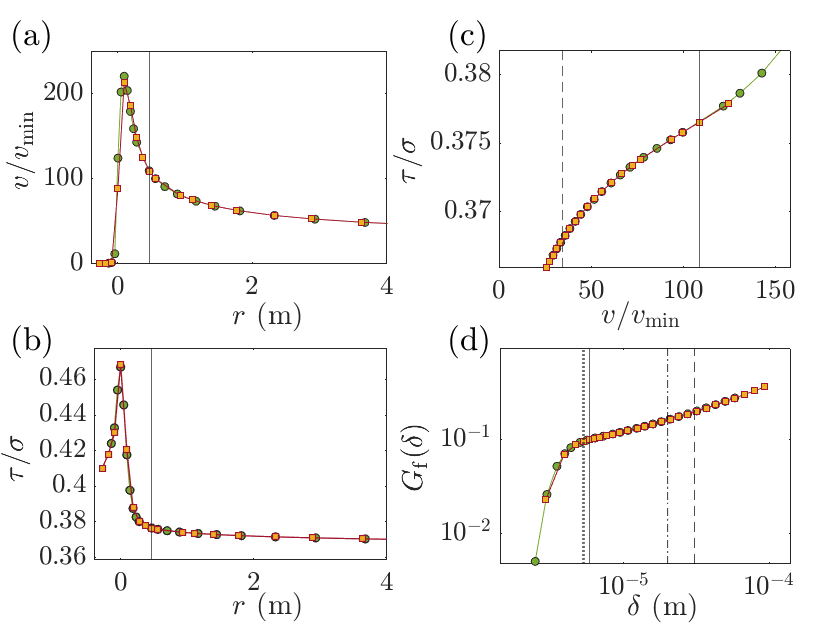}
\vspace{-0.5cm}
\caption{Convergence test for $c_{\rm r}/c_{_{\rm R}}\!\simeq\!0.27$. (a) The slip rate snapshots, (b) the shear stress snapshots, (c) the snapshots of $\tau(v)$,  and (d) $G_{\rm f}(\delta)$ for the two resolutions discussed in the text. Yellow squares denote the lower resolution results ($\Delta{x}\!=\!0.093$ m) and green circles mark the higher resolution results ($\Delta{x}\!=\!0.0465$ m). The vertical solid line in all panels denotes the values of $\delta_{\rm c}$ obtained from higher resolution simulations. In panel (d), the dotted line marks the value of $\delta_{\rm c}$ in the lower resolution simulations. In panels (c) and (d), the dashed and dot-dashed lines denote the end of the power-law scaling range for higher and lower resolution, respectively.}
\label{fig:f6sm}
\end{figure}
\begin{figure}[htbp]
\centering
\includegraphics[width=0.38\textwidth]{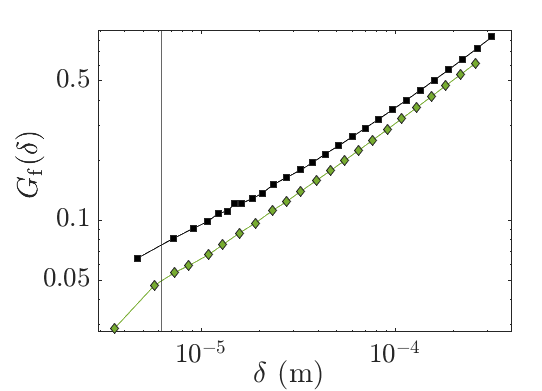}
\vspace{-0.25cm}
\caption{The low numerical resolution (cf.~Fig.~\ref{fig:f6sm} and text for definition) is not sufficient to resolve the edge of the cohesive zone very close to $c_{_{\rm R}}$. We present the breakdown energy $G_{\rm f}(\delta)$ for $c_{\rm r}\!\simeq\!0.93c_{_{\rm R}}$ (green diamonds) and $c_{\rm r}\!\simeq\!0.99c_{_{\rm R}}$ (black squares). For the former, $\delta_{\rm c}$ is resolved and is marked by the vertical solid line. For the latter, $\delta_{\rm c}$ is not resolved (at best, the results offer an upper bound on $\delta_{\rm c}$). This issue is rectified by using the higher resolution, as demonstrated in Fig.~\ref{fig:f8sm}.}
\label{fig:f7sm}
\end{figure}
\begin{figure}[htbp]
\includegraphics[width=0.5\textwidth]{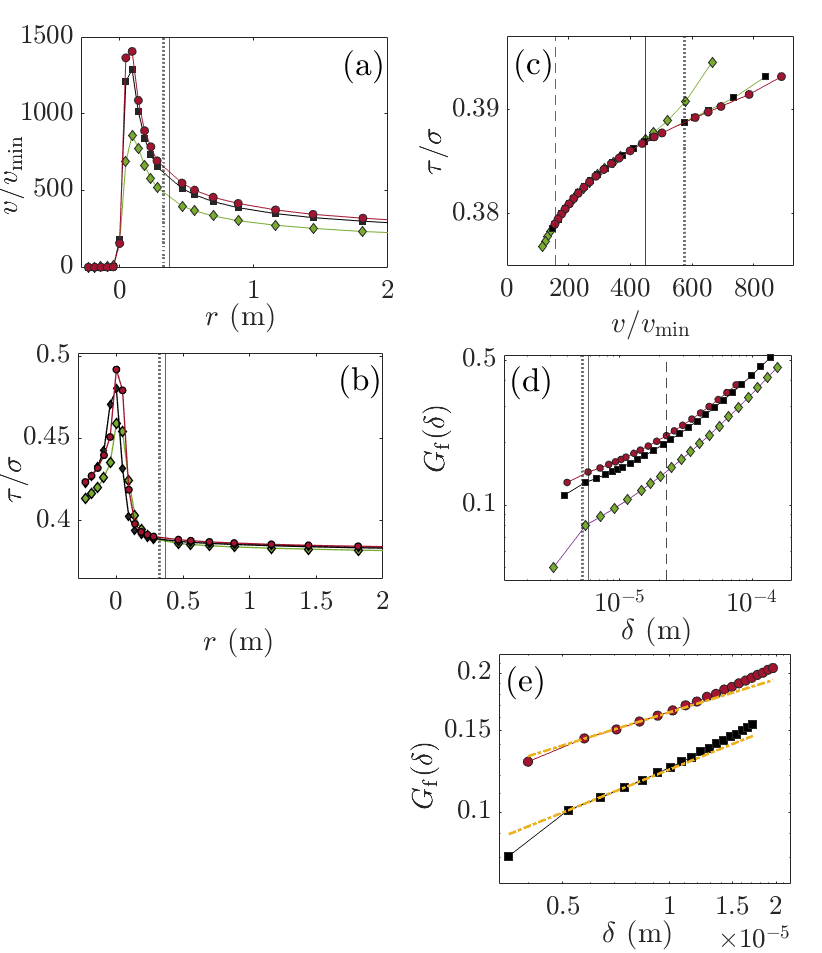}
\centering
\vspace{-0.35cm}
\caption{The high numerical resolution (cf.~Fig.~\ref{fig:f6sm} and text for definition) is sufficient to resolve the edge of the cohesive zone very close to $c_{_{\rm R}}$ and above it. (a) The slip rate snapshots for $c_{\rm r}\!\simeq\!0.93c_{_{\rm R}}$ (green diamonds), $c_{\rm r}\!\simeq\!0.99c_{_{\rm R}}$ (black squares), and $c_{\rm r}\!\simeq\!1.014c_{_{\rm R}}\!>\!c_{_{\rm R}}$ (brown circles), all symbols as in Fig.~4(inset) in the manuscript. (b) the corresponding shear stress snapshots. (c) the corresponding snapshots of $\tau(v)$. (d) The corresponding $G_{\rm f}(\delta)$, demonstrating that $\delta_{\rm c}$ (and hence $\ell_{\rm c}$) is resolved at all $3$ propagation speeds, as is further highlighted in panel (e). $\delta_{\rm c}$ for $c_{\rm r}\!\simeq\!0.93c_{_{\rm R}}$ is marked in panel (d) by the vertical solid line, while for both $c_{\rm r}\!\simeq\!0.99c_{_{\rm R}}$ and $c_{\rm r}\!\simeq\!1.014c_{_{\rm R}}$ it is marked by the vertical dotted line (its value is the same for the two speeds, within the current uncertainty). The upper limit of the power-law scaling range of $G_{\rm f}(\delta)$ for $c_{\rm r}\!\simeq\!0.93c_{_{\rm R}}$ is marked in panel (d) by the vertical dashed line. The corresponding lines are superposed on panel (c). Finally, the corresponding values of $\ell_{\rm c}$ are marked in panels (a)-(b). (e) A zoom-in on the two upper datasets in panel (d), highlighting that the first data point in each slightly deviates from the subsequent data points, which allows to estimate $\delta_{\rm c}$.}
\label{fig:f8sm}
\end{figure}

To demonstrate this, we compare results for rupture propagation in the low sub-Rayleigh regime ($c_{\rm r}\!\simeq\!0.27c_{_{\rm R}}$, cf.~Fig.~1 in the manuscript) and close to $c_{_{\rm R}}$ ($c_{\rm r}\!\simeq\!0.93c_{_{\rm R}}$, cf.~Fig.~2 in the manuscript) for two numerical resolutions $\Delta{x}\=0.093$ m (and correspondingly, $\Delta{t}\=6.79\!\times\!10^{-6}$ sec) and $\Delta{x}\=0.0465$ m (and correspondingly, $\Delta{t}\=3.395\!\times\!10^{-6}$ sec). First, we demonstrate in Fig.~\ref{fig:f6sm} that our simulations are fully converged in the low sub-Rayleigh regime. Such a quality of convergence persists up to rupture propagation speeds of about $0.9c_{\rm s}$ or so. Second, for higher propagation speeds, the lower spatial resolution is not sufficient to resolve the edge of the cohesion zone, as demonstrated in Fig.~\ref{fig:f7sm}, where we present results for $c_{\rm r}\!\simeq\!0.93c_{_{\rm R}}$ and $c_{\rm r}\!\simeq\!0.99c_{_{\rm R}}$ using the lower resolution. Finally, this issue is rectified by employing the higher resolution, as demonstrated in Fig.~\ref{fig:f8sm}.

Additional refinement of the spatial resolution is computationally prohibitive for the problem at hand, in light of the physical questions that are addressed. Yet, the above analysis demonstrated the robustness and numerical convergence of our main computational findings, which agree with the theoretical predictions, using the higher resolution scheme. Consequently, all the results presented in the manuscript and in the previous sections in this document correspond to the higher resolution of the two discussed.

\newpage


\begin{thebibliography}{85}%
\makeatletter
\providecommand \@ifxundefined [1]{%
 \@ifx{#1\undefined}
}%
\providecommand \@ifnum [1]{%
 \ifnum #1\expandafter \@firstoftwo
 \else \expandafter \@secondoftwo
 \fi
}%
\providecommand \@ifx [1]{%
 \ifx #1\expandafter \@firstoftwo
 \else \expandafter \@secondoftwo
 \fi
}%
\providecommand \natexlab [1]{#1}%
\providecommand \enquote  [1]{``#1''}%
\providecommand \bibnamefont  [1]{#1}%
\providecommand \bibfnamefont [1]{#1}%
\providecommand \citenamefont [1]{#1}%
\providecommand \href@noop [0]{\@secondoftwo}%
\providecommand \href [0]{\begingroup \@sanitize@url \@href}%
\providecommand \@href[1]{\@@startlink{#1}\@@href}%
\providecommand \@@href[1]{\endgroup#1\@@endlink}%
\providecommand \@sanitize@url [0]{\catcode `\\12\catcode `\$12\catcode
  `\&12\catcode `\#12\catcode `\^12\catcode `\_12\catcode `\%12\relax}%
\providecommand \@@startlink[1]{}%
\providecommand \@@endlink[0]{}%
\providecommand \url  [0]{\begingroup\@sanitize@url \@url }%
\providecommand \@url [1]{\endgroup\@href {#1}{\urlprefix }}%
\providecommand \urlprefix  [0]{URL }%
\providecommand \Eprint [0]{\href }%
\providecommand \doibase [0]{https://doi.org/}%
\providecommand \selectlanguage [0]{\@gobble}%
\providecommand \bibinfo  [0]{\@secondoftwo}%
\providecommand \bibfield  [0]{\@secondoftwo}%
\providecommand \translation [1]{[#1]}%
\providecommand \BibitemOpen [0]{}%
\providecommand \bibitemStop [0]{}%
\providecommand \bibitemNoStop [0]{.\EOS\space}%
\providecommand \EOS [0]{\spacefactor3000\relax}%
\providecommand \BibitemShut  [1]{\csname bibitem#1\endcsname}%
\let\auto@bib@innerbib\@empty
\bibitem [{\citenamefont {Scholz}(2002)}]{Scholz2002}%
  \BibitemOpen
  \bibfield  {author} {\bibinfo {author} {\bibfnamefont {C.~H.}\ \bibnamefont
  {Scholz}},\ }\href {https://doi.org/10.1017/9781316681473} {\emph {\bibinfo
  {title} {{The Mechanics of Earthquakes and Faulting}}}}\ (\bibinfo
  {publisher} {Cambridge university press},\ \bibinfo {year}
  {2002})\BibitemShut {NoStop}%
\bibitem [{\citenamefont {Bao}\ \emph {et~al.}(2022)\citenamefont {Bao},
  \citenamefont {Xu}, \citenamefont {Meng}, \citenamefont {Ampuero},
  \citenamefont {Gao},\ and\ \citenamefont {Zhang}}]{bao2022global}%
  \BibitemOpen
  \bibfield  {author} {\bibinfo {author} {\bibfnamefont {H.}~\bibnamefont
  {Bao}}, \bibinfo {author} {\bibfnamefont {L.}~\bibnamefont {Xu}}, \bibinfo
  {author} {\bibfnamefont {L.}~\bibnamefont {Meng}}, \bibinfo {author}
  {\bibfnamefont {J.-P.}\ \bibnamefont {Ampuero}}, \bibinfo {author}
  {\bibfnamefont {L.}~\bibnamefont {Gao}},\ and\ \bibinfo {author}
  {\bibfnamefont {H.}~\bibnamefont {Zhang}},\ }\bibfield  {title} {\bibinfo
  {title} {Global frequency of oceanic and continental supershear
  earthquakes},\ }\href {https://www.nature.com/articles/s41561-022-01055-5}
  {\bibfield  {journal} {\bibinfo  {journal} {Nature Geoscience}\ }\textbf
  {\bibinfo {volume} {15}},\ \bibinfo {pages} {942} (\bibinfo {year}
  {2022})}\BibitemShut {NoStop}%
\bibitem [{\citenamefont {Elbanna}\ \emph {et~al.}(2025)\citenamefont
  {Elbanna}, \citenamefont {Abdelmeguid}, \citenamefont {Asimaki},
  \citenamefont {Tainpakdipat}, \citenamefont {Lavrentadis}, \citenamefont
  {Rosakis},\ and\ \citenamefont {Ben-Zion}}]{elbanna2025supershear}%
  \BibitemOpen
  \bibfield  {author} {\bibinfo {author} {\bibfnamefont {A.}~\bibnamefont
  {Elbanna}}, \bibinfo {author} {\bibfnamefont {M.}~\bibnamefont
  {Abdelmeguid}}, \bibinfo {author} {\bibfnamefont {D.}~\bibnamefont
  {Asimaki}}, \bibinfo {author} {\bibfnamefont {N.}~\bibnamefont
  {Tainpakdipat}}, \bibinfo {author} {\bibfnamefont {G.}~\bibnamefont
  {Lavrentadis}}, \bibinfo {author} {\bibfnamefont {A.}~\bibnamefont
  {Rosakis}},\ and\ \bibinfo {author} {\bibfnamefont {Y.}~\bibnamefont
  {Ben-Zion}},\ }\bibfield  {title} {\bibinfo {title} {{Supershear Earthquakes:
  Their Occurrence and Importance for Seismic Hazard, Early Warning, and Design
  Standards}},\ }\href {https://doi.org/10.1785/0220250118} {\bibfield
  {journal} {\bibinfo  {journal} {Seismological Research Letters}\ }\textbf
  {\bibinfo {volume} {96}},\ \bibinfo {pages} {3319} (\bibinfo {year}
  {2025})}\BibitemShut {NoStop}%
\bibitem [{\citenamefont {Rosakis}(2002)}]{Rosakis2002}%
  \BibitemOpen
  \bibfield  {author} {\bibinfo {author} {\bibfnamefont {A.~J.}\ \bibnamefont
  {Rosakis}},\ }\bibfield  {title} {\bibinfo {title} {{Intersonic shear cracks
  and fault ruptures}},\ }\href {https://doi.org/10.1080/00018730210122328}
  {\bibfield  {journal} {\bibinfo  {journal} {Adv. Phys.}\ }\textbf {\bibinfo
  {volume} {51}},\ \bibinfo {pages} {1189} (\bibinfo {year}
  {2002})}\BibitemShut {NoStop}%
\bibitem [{\citenamefont {Bizzarri}(2019)}]{bizzarri2019mechanics}%
  \BibitemOpen
  \bibfield  {author} {\bibinfo {author} {\bibfnamefont {A.}~\bibnamefont
  {Bizzarri}},\ }\bibfield  {title} {\bibinfo {title} {The mechanics of
  supershear earthquake ruptures},\ }\href
  {https://doi.org/10.3254/978-1-61499-979-9-1} {\bibfield  {journal} {\bibinfo
   {journal} {Mechanics of Earthquake Faulting}\ ,\ \bibinfo {pages} {1}}
  (\bibinfo {year} {2019})}\BibitemShut {NoStop}%
\bibitem [{\citenamefont {Bernard}\ and\ \citenamefont
  {Baumont}(2005)}]{bernard2005shear}%
  \BibitemOpen
  \bibfield  {author} {\bibinfo {author} {\bibfnamefont {P.}~\bibnamefont
  {Bernard}}\ and\ \bibinfo {author} {\bibfnamefont {D.}~\bibnamefont
  {Baumont}},\ }\bibfield  {title} {\bibinfo {title} {Shear mach wave
  characterization for kinematic fault rupture models with constant supershear
  rupture velocity},\ }\href
  {https://academic.oup.com/gji/article/162/2/431/552341} {\bibfield  {journal}
  {\bibinfo  {journal} {Geophysical Journal International}\ }\textbf {\bibinfo
  {volume} {162}},\ \bibinfo {pages} {431} (\bibinfo {year}
  {2005})}\BibitemShut {NoStop}%
\bibitem [{\citenamefont {Dunham}\ and\ \citenamefont
  {Archuleta}(2005)}]{dunham2005near}%
  \BibitemOpen
  \bibfield  {author} {\bibinfo {author} {\bibfnamefont {E.~M.}\ \bibnamefont
  {Dunham}}\ and\ \bibinfo {author} {\bibfnamefont {R.~J.}\ \bibnamefont
  {Archuleta}},\ }\bibfield  {title} {\bibinfo {title} {Near-source ground
  motion from steady state dynamic rupture pulses},\ }\href
  {https://agupubs.onlinelibrary.wiley.com/doi/full/10.1029/2004GL021793}
  {\bibfield  {journal} {\bibinfo  {journal} {Geophysical Research Letters}\
  }\textbf {\bibinfo {volume} {32}} (\bibinfo {year} {2005})}\BibitemShut
  {NoStop}%
\bibitem [{\citenamefont {Bhat}\ \emph {et~al.}(2007)\citenamefont {Bhat},
  \citenamefont {Dmowska}, \citenamefont {King}, \citenamefont {Klinger},\ and\
  \citenamefont {Rice}}]{bhat2007off}%
  \BibitemOpen
  \bibfield  {author} {\bibinfo {author} {\bibfnamefont {H.~S.}\ \bibnamefont
  {Bhat}}, \bibinfo {author} {\bibfnamefont {R.}~\bibnamefont {Dmowska}},
  \bibinfo {author} {\bibfnamefont {G.~C.}\ \bibnamefont {King}}, \bibinfo
  {author} {\bibfnamefont {Y.}~\bibnamefont {Klinger}},\ and\ \bibinfo {author}
  {\bibfnamefont {J.~R.}\ \bibnamefont {Rice}},\ }\bibfield  {title} {\bibinfo
  {title} {{Off-fault damage patterns due to supershear ruptures with
  application to the 2001 Mw 8.1 Kokoxili (Kunlun) Tibet earthquake}},\ }\href
  {https://agupubs.onlinelibrary.wiley.com/doi/10.1029/2006JB004425} {\bibfield
   {journal} {\bibinfo  {journal} {Journal of Geophysical Research: Solid
  Earth}\ }\textbf {\bibinfo {volume} {112}} (\bibinfo {year}
  {2007})}\BibitemShut {NoStop}%
\bibitem [{\citenamefont {Freund}(1979)}]{Freund1979}%
  \BibitemOpen
  \bibfield  {author} {\bibinfo {author} {\bibfnamefont {L.~B.}\ \bibnamefont
  {Freund}},\ }\bibfield  {title} {\bibinfo {title} {{The mechanics of dynamic
  shear crack propagation}},\ }\href {https://doi.org/10.1029/JB084iB05p02199}
  {\bibfield  {journal} {\bibinfo  {journal} {J. Geophys. Res.}\ }\textbf
  {\bibinfo {volume} {84}},\ \bibinfo {pages} {2199} (\bibinfo {year}
  {1979})}\BibitemShut {NoStop}%
\bibitem [{\citenamefont {Freund}(1998)}]{Freund1998}%
  \BibitemOpen
  \bibfield  {author} {\bibinfo {author} {\bibfnamefont {L.~B.}\ \bibnamefont
  {Freund}},\ }\href {https://doi.org/10.1017/CBO9780511546761} {\emph
  {\bibinfo {title} {{Dynamic Fracture Mechanics}}}}\ (\bibinfo  {publisher}
  {Cambridge university press},\ \bibinfo {address} {Cambridge},\ \bibinfo
  {year} {1998})\BibitemShut {NoStop}%
\bibitem [{\citenamefont {Broberg}(1999)}]{Broberg1999}%
  \BibitemOpen
  \bibfield  {author} {\bibinfo {author} {\bibfnamefont {K.~B.}\ \bibnamefont
  {Broberg}},\ }\href
  {https://www.elsevier.com/books/cracks-and-fracture/broberg/978-0-12-134130-5}
  {\emph {\bibinfo {title} {{Cracks and fracture}}}}\ (\bibinfo  {publisher}
  {Academic Press},\ \bibinfo {year} {1999})\BibitemShut {NoStop}%
\bibitem [{\citenamefont {Burridge}(1973)}]{burridge1973admissible}%
  \BibitemOpen
  \bibfield  {author} {\bibinfo {author} {\bibfnamefont {R.}~\bibnamefont
  {Burridge}},\ }\bibfield  {title} {\bibinfo {title} {Admissible speeds for
  plane-strain self-similar shear cracks with friction but lacking cohesion},\
  }\href {https://doi.org/10.1111/j.1365-246X.1973.tb00608.x} {\bibfield
  {journal} {\bibinfo  {journal} {Geophysical Journal International}\ }\textbf
  {\bibinfo {volume} {35}},\ \bibinfo {pages} {439} (\bibinfo {year}
  {1973})}\BibitemShut {NoStop}%
\bibitem [{\citenamefont {Andrews}(1976)}]{andrews1976rupture}%
  \BibitemOpen
  \bibfield  {author} {\bibinfo {author} {\bibfnamefont {D.}~\bibnamefont
  {Andrews}},\ }\bibfield  {title} {\bibinfo {title} {Rupture velocity of plane
  strain shear cracks},\ }\href {https://doi.org/10.1029/JB081i032p05679}
  {\bibfield  {journal} {\bibinfo  {journal} {Journal of Geophysical Research}\
  }\textbf {\bibinfo {volume} {81}},\ \bibinfo {pages} {5679} (\bibinfo {year}
  {1976})}\BibitemShut {NoStop}%
\bibitem [{\citenamefont {Burridge}\ \emph {et~al.}(1979)\citenamefont
  {Burridge}, \citenamefont {Conn},\ and\ \citenamefont
  {Freund}}]{burridge1979stability}%
  \BibitemOpen
  \bibfield  {author} {\bibinfo {author} {\bibfnamefont {R.}~\bibnamefont
  {Burridge}}, \bibinfo {author} {\bibfnamefont {G.}~\bibnamefont {Conn}},\
  and\ \bibinfo {author} {\bibfnamefont {L.~B.}\ \bibnamefont {Freund}},\
  }\bibfield  {title} {\bibinfo {title} {{The stability of a rapid mode II
  shear crack with finite cohesive traction}},\ }\href
  {https://doi.org/10.1029/JB084iB05p02210} {\bibfield  {journal} {\bibinfo
  {journal} {Journal of Geophysical Research: Solid Earth}\ }\textbf {\bibinfo
  {volume} {84}},\ \bibinfo {pages} {2210} (\bibinfo {year}
  {1979})}\BibitemShut {NoStop}%
\bibitem [{\citenamefont {Gabriel}\ \emph {et~al.}(2012)\citenamefont
  {Gabriel}, \citenamefont {Ampuero}, \citenamefont {Dalguer},\ and\
  \citenamefont {Mai}}]{Gabriel2012}%
  \BibitemOpen
  \bibfield  {author} {\bibinfo {author} {\bibfnamefont {A.-A.}\ \bibnamefont
  {Gabriel}}, \bibinfo {author} {\bibfnamefont {J.-P.}\ \bibnamefont
  {Ampuero}}, \bibinfo {author} {\bibfnamefont {L.~A.}\ \bibnamefont
  {Dalguer}},\ and\ \bibinfo {author} {\bibfnamefont {P.~M.}\ \bibnamefont
  {Mai}},\ }\bibfield  {title} {\bibinfo {title} {{The transition of dynamic
  rupture styles in elastic media under velocity-weakening friction}},\ }\href
  {https://doi.org/10.1029/2012JB009468} {\bibfield  {journal} {\bibinfo
  {journal} {J. Geophys. Res. Solid Earth}\ }\textbf {\bibinfo {volume}
  {117}},\ \bibinfo {pages} {B09311} (\bibinfo {year} {2012})}\BibitemShut
  {NoStop}%
\bibitem [{\citenamefont {Svetlizky}\ \emph {et~al.}(2016)\citenamefont
  {Svetlizky}, \citenamefont {{Pino Mu{\~{n}}oz}}, \citenamefont {Radiguet},
  \citenamefont {Kammer}, \citenamefont {Molinari},\ and\ \citenamefont
  {Fineberg}}]{Svetlizky2016}%
  \BibitemOpen
  \bibfield  {author} {\bibinfo {author} {\bibfnamefont {I.}~\bibnamefont
  {Svetlizky}}, \bibinfo {author} {\bibfnamefont {D.}~\bibnamefont {{Pino
  Mu{\~{n}}oz}}}, \bibinfo {author} {\bibfnamefont {M.}~\bibnamefont
  {Radiguet}}, \bibinfo {author} {\bibfnamefont {D.~S.}\ \bibnamefont
  {Kammer}}, \bibinfo {author} {\bibfnamefont {J.-F.}\ \bibnamefont
  {Molinari}},\ and\ \bibinfo {author} {\bibfnamefont {J.}~\bibnamefont
  {Fineberg}},\ }\bibfield  {title} {\bibinfo {title} {{Properties of the shear
  stress peak radiated ahead of rapidly accelerating rupture fronts that
  mediate frictional slip}},\ }\href {https://doi.org/10.1073/pnas.1517545113}
  {\bibfield  {journal} {\bibinfo  {journal} {Proc. Natl. Acad. Sci.}\ }\textbf
  {\bibinfo {volume} {113}},\ \bibinfo {pages} {542} (\bibinfo {year}
  {2016})}\BibitemShut {NoStop}%
\bibitem [{\citenamefont {Dunham}(2007)}]{dunham2007conditions}%
  \BibitemOpen
  \bibfield  {author} {\bibinfo {author} {\bibfnamefont {E.~M.}\ \bibnamefont
  {Dunham}},\ }\bibfield  {title} {\bibinfo {title} {Conditions governing the
  occurrence of supershear ruptures under slip-weakening friction},\ }\href
  {https://doi.org/10.1029/2006JB004717} {\bibfield  {journal} {\bibinfo
  {journal} {Journal of Geophysical Research: Solid Earth}\ }\textbf {\bibinfo
  {volume} {112}},\ \bibinfo {pages} {B07302} (\bibinfo {year}
  {2007})}\BibitemShut {NoStop}%
\bibitem [{\citenamefont {Fukuyama}\ and\ \citenamefont
  {Olsen}(2002)}]{fukuyama2002condition}%
  \BibitemOpen
  \bibfield  {author} {\bibinfo {author} {\bibfnamefont {E.}~\bibnamefont
  {Fukuyama}}\ and\ \bibinfo {author} {\bibfnamefont {K.~B.}\ \bibnamefont
  {Olsen}},\ }\bibfield  {title} {\bibinfo {title} {A condition for super-shear
  rupture propagation in a heterogeneous stress field},\ }\href
  {https://link.springer.com/article/10.1007/s00024-002-8722-y} {\bibfield
  {journal} {\bibinfo  {journal} {Pure and Applied Geophysics}\ }\textbf
  {\bibinfo {volume} {159}},\ \bibinfo {pages} {2047} (\bibinfo {year}
  {2002})}\BibitemShut {NoStop}%
\bibitem [{\citenamefont {Dunham}\ \emph {et~al.}(2003)\citenamefont {Dunham},
  \citenamefont {Favreau},\ and\ \citenamefont
  {Carlson}}]{dunham2003supershear}%
  \BibitemOpen
  \bibfield  {author} {\bibinfo {author} {\bibfnamefont {E.~M.}\ \bibnamefont
  {Dunham}}, \bibinfo {author} {\bibfnamefont {P.}~\bibnamefont {Favreau}},\
  and\ \bibinfo {author} {\bibfnamefont {J.}~\bibnamefont {Carlson}},\
  }\bibfield  {title} {\bibinfo {title} {A supershear transition mechanism for
  cracks},\ }\href {https://www.science.org/doi/10.1126/science.1080650}
  {\bibfield  {journal} {\bibinfo  {journal} {Science}\ }\textbf {\bibinfo
  {volume} {299}},\ \bibinfo {pages} {1557} (\bibinfo {year}
  {2003})}\BibitemShut {NoStop}%
\bibitem [{\citenamefont {Liu}\ and\ \citenamefont {Lapusta}(2008)}]{Liu2008}%
  \BibitemOpen
  \bibfield  {author} {\bibinfo {author} {\bibfnamefont {Y.}~\bibnamefont
  {Liu}}\ and\ \bibinfo {author} {\bibfnamefont {N.}~\bibnamefont {Lapusta}},\
  }\bibfield  {title} {\bibinfo {title} {{Transition of mode II cracks from
  sub-Rayleigh to intersonic speeds in the presence of favorable
  heterogeneity}},\ }\href {https://doi.org/10.1016/j.jmps.2007.06.005}
  {\bibfield  {journal} {\bibinfo  {journal} {J. Mech. Phys. Solids}\ }\textbf
  {\bibinfo {volume} {56}},\ \bibinfo {pages} {25} (\bibinfo {year}
  {2008})}\BibitemShut {NoStop}%
\bibitem [{\citenamefont {Weng}\ \emph {et~al.}(2015)\citenamefont {Weng},
  \citenamefont {Huang},\ and\ \citenamefont {Yang}}]{weng2015barrier}%
  \BibitemOpen
  \bibfield  {author} {\bibinfo {author} {\bibfnamefont {H.}~\bibnamefont
  {Weng}}, \bibinfo {author} {\bibfnamefont {J.}~\bibnamefont {Huang}},\ and\
  \bibinfo {author} {\bibfnamefont {H.}~\bibnamefont {Yang}},\ }\bibfield
  {title} {\bibinfo {title} {Barrier-induced supershear ruptures on a
  slip-weakening fault},\ }\href {https://doi.org/10.1002/2015GL064281}
  {\bibfield  {journal} {\bibinfo  {journal} {Geophysical Research Letters}\
  }\textbf {\bibinfo {volume} {42}},\ \bibinfo {pages} {4824} (\bibinfo {year}
  {2015})}\BibitemShut {NoStop}%
\bibitem [{\citenamefont {Barras}\ \emph {et~al.}(2017)\citenamefont {Barras},
  \citenamefont {Geubelle},\ and\ \citenamefont
  {Molinari}}]{barras2017interplay}%
  \BibitemOpen
  \bibfield  {author} {\bibinfo {author} {\bibfnamefont {F.}~\bibnamefont
  {Barras}}, \bibinfo {author} {\bibfnamefont {P.~H.}\ \bibnamefont
  {Geubelle}},\ and\ \bibinfo {author} {\bibfnamefont {J.-F.}\ \bibnamefont
  {Molinari}},\ }\bibfield  {title} {\bibinfo {title} {Interplay between
  process zone and material heterogeneities for dynamic cracks},\ }\href
  {https://journals.aps.org/prl/abstract/10.1103/PhysRevLett.119.144101}
  {\bibfield  {journal} {\bibinfo  {journal} {Physical Review Letters}\
  }\textbf {\bibinfo {volume} {119}},\ \bibinfo {pages} {144101} (\bibinfo
  {year} {2017})}\BibitemShut {NoStop}%
\bibitem [{\citenamefont {Ryan}\ and\ \citenamefont
  {Oglesby}(2014)}]{ryan2014dynamically}%
  \BibitemOpen
  \bibfield  {author} {\bibinfo {author} {\bibfnamefont {K.~J.}\ \bibnamefont
  {Ryan}}\ and\ \bibinfo {author} {\bibfnamefont {D.~D.}\ \bibnamefont
  {Oglesby}},\ }\bibfield  {title} {\bibinfo {title} {Dynamically modeling
  fault step overs using various friction laws},\ }\href
  {https://doi.org/10.1002/2014JB011151} {\bibfield  {journal} {\bibinfo
  {journal} {Journal of Geophysical Research: Solid Earth}\ }\textbf {\bibinfo
  {volume} {119}},\ \bibinfo {pages} {5814} (\bibinfo {year}
  {2014})}\BibitemShut {NoStop}%
\bibitem [{\citenamefont {Hu}\ \emph {et~al.}(2016)\citenamefont {Hu},
  \citenamefont {Xu}, \citenamefont {Zhang},\ and\ \citenamefont
  {Chen}}]{hu2016supershear}%
  \BibitemOpen
  \bibfield  {author} {\bibinfo {author} {\bibfnamefont {F.}~\bibnamefont
  {Hu}}, \bibinfo {author} {\bibfnamefont {J.}~\bibnamefont {Xu}}, \bibinfo
  {author} {\bibfnamefont {Z.}~\bibnamefont {Zhang}},\ and\ \bibinfo {author}
  {\bibfnamefont {X.}~\bibnamefont {Chen}},\ }\bibfield  {title} {\bibinfo
  {title} {Supershear transition mechanism induced by step over geometry},\
  }\href {https://doi.org/10.1002/2016JB013333} {\bibfield  {journal} {\bibinfo
   {journal} {Journal of Geophysical Research: Solid Earth}\ }\textbf {\bibinfo
  {volume} {121}},\ \bibinfo {pages} {8738} (\bibinfo {year}
  {2016})}\BibitemShut {NoStop}%
\bibitem [{\citenamefont {Kehoe}\ and\ \citenamefont
  {Kiser}(2020)}]{kehoe2020evidence}%
  \BibitemOpen
  \bibfield  {author} {\bibinfo {author} {\bibfnamefont {H.}~\bibnamefont
  {Kehoe}}\ and\ \bibinfo {author} {\bibfnamefont {E.}~\bibnamefont {Kiser}},\
  }\bibfield  {title} {\bibinfo {title} {Evidence of a supershear transition
  across a fault stepover},\ }\href {https://doi.org/10.1029/2020GL087400}
  {\bibfield  {journal} {\bibinfo  {journal} {Geophysical Research Letters}\
  }\textbf {\bibinfo {volume} {47}},\ \bibinfo {pages} {e2020GL087400}
  (\bibinfo {year} {2020})}\BibitemShut {NoStop}%
\bibitem [{\citenamefont {Bruhat}\ \emph {et~al.}(2016)\citenamefont {Bruhat},
  \citenamefont {Fang},\ and\ \citenamefont {Dunham}}]{Bruhat2016}%
  \BibitemOpen
  \bibfield  {author} {\bibinfo {author} {\bibfnamefont {L.}~\bibnamefont
  {Bruhat}}, \bibinfo {author} {\bibfnamefont {Z.}~\bibnamefont {Fang}},\ and\
  \bibinfo {author} {\bibfnamefont {E.~M.}\ \bibnamefont {Dunham}},\ }\bibfield
   {title} {\bibinfo {title} {{Rupture complexity and the supershear transition
  on rough faults}},\ }\href {https://doi.org/10.1002/2015JB012512} {\bibfield
  {journal} {\bibinfo  {journal} {J. Geophys. Res. Solid Earth}\ }\textbf
  {\bibinfo {volume} {121}},\ \bibinfo {pages} {210} (\bibinfo {year}
  {2016})}\BibitemShut {NoStop}%
\bibitem [{\citenamefont {Huang}\ \emph {et~al.}(2016)\citenamefont {Huang},
  \citenamefont {Ampuero},\ and\ \citenamefont {Helmberger}}]{Huang2016a}%
  \BibitemOpen
  \bibfield  {author} {\bibinfo {author} {\bibfnamefont {Y.}~\bibnamefont
  {Huang}}, \bibinfo {author} {\bibfnamefont {J.-P.}\ \bibnamefont {Ampuero}},\
  and\ \bibinfo {author} {\bibfnamefont {D.~V.}\ \bibnamefont {Helmberger}},\
  }\bibfield  {title} {\bibinfo {title} {{The potential for supershear
  earthquakes in damaged fault zones – theory and observations}},\ }\href
  {https://doi.org/10.1016/j.epsl.2015.10.046} {\bibfield  {journal} {\bibinfo
  {journal} {Earth Planet. Sci. Lett.}\ }\textbf {\bibinfo {volume} {433}},\
  \bibinfo {pages} {109} (\bibinfo {year} {2016})}\BibitemShut {NoStop}%
\bibitem [{\citenamefont {Kaneko}\ and\ \citenamefont
  {Lapusta}(2010)}]{kaneko2010supershear}%
  \BibitemOpen
  \bibfield  {author} {\bibinfo {author} {\bibfnamefont {Y.}~\bibnamefont
  {Kaneko}}\ and\ \bibinfo {author} {\bibfnamefont {N.}~\bibnamefont
  {Lapusta}},\ }\bibfield  {title} {\bibinfo {title} {{Supershear transition
  due to a free surface in 3-D simulations of spontaneous dynamic rupture on
  vertical strike-slip faults}},\ }\href
  {https://www.sciencedirect.com/science/article/pii/S0040195110002684}
  {\bibfield  {journal} {\bibinfo  {journal} {Tectonophysics}\ }\textbf
  {\bibinfo {volume} {493}},\ \bibinfo {pages} {272} (\bibinfo {year}
  {2010})}\BibitemShut {NoStop}%
\bibitem [{\citenamefont {Weng}\ and\ \citenamefont
  {Ampuero}(2020)}]{weng2020continuum}%
  \BibitemOpen
  \bibfield  {author} {\bibinfo {author} {\bibfnamefont {H.}~\bibnamefont
  {Weng}}\ and\ \bibinfo {author} {\bibfnamefont {J.-P.}\ \bibnamefont
  {Ampuero}},\ }\bibfield  {title} {\bibinfo {title} {Continuum of earthquake
  rupture speeds enabled by oblique slip},\ }\href
  {https://doi.org/10.1038/s41561-020-00654-4} {\bibfield  {journal} {\bibinfo
  {journal} {Nature Geoscience}\ }\textbf {\bibinfo {volume} {13}},\ \bibinfo
  {pages} {817} (\bibinfo {year} {2020})}\BibitemShut {NoStop}%
\bibitem [{\citenamefont {Rosakis}\ \emph {et~al.}(1999)\citenamefont
  {Rosakis}, \citenamefont {Samudrala},\ and\ \citenamefont
  {Coker}}]{rosakis1999cracks}%
  \BibitemOpen
  \bibfield  {author} {\bibinfo {author} {\bibfnamefont {A.~J.}\ \bibnamefont
  {Rosakis}}, \bibinfo {author} {\bibfnamefont {O.}~\bibnamefont {Samudrala}},\
  and\ \bibinfo {author} {\bibfnamefont {D.}~\bibnamefont {Coker}},\ }\bibfield
   {title} {\bibinfo {title} {Cracks faster than the shear wave speed},\ }\href
  {https://www.science.org/doi/10.1126/science.284.5418.1337} {\bibfield
  {journal} {\bibinfo  {journal} {Science}\ }\textbf {\bibinfo {volume}
  {284}},\ \bibinfo {pages} {1337} (\bibinfo {year} {1999})}\BibitemShut
  {NoStop}%
\bibitem [{\citenamefont {Xia}\ \emph {et~al.}(2004)\citenamefont {Xia},
  \citenamefont {Rosakis},\ and\ \citenamefont {Kanamori}}]{xia2004laboratory}%
  \BibitemOpen
  \bibfield  {author} {\bibinfo {author} {\bibfnamefont {K.}~\bibnamefont
  {Xia}}, \bibinfo {author} {\bibfnamefont {A.~J.}\ \bibnamefont {Rosakis}},\
  and\ \bibinfo {author} {\bibfnamefont {H.}~\bibnamefont {Kanamori}},\
  }\bibfield  {title} {\bibinfo {title} {{Laboratory earthquakes: The
  sub-Rayleigh-to-supershear rupture transition}},\ }\href
  {http://science.org/doi/10.1126/science.1094022} {\bibfield  {journal}
  {\bibinfo  {journal} {Science}\ }\textbf {\bibinfo {volume} {303}},\ \bibinfo
  {pages} {1859} (\bibinfo {year} {2004})}\BibitemShut {NoStop}%
\bibitem [{\citenamefont {Ben-David}\ \emph {et~al.}(2010)\citenamefont
  {Ben-David}, \citenamefont {Cohen},\ and\ \citenamefont
  {Fineberg}}]{Ben-David2010a}%
  \BibitemOpen
  \bibfield  {author} {\bibinfo {author} {\bibfnamefont {O.}~\bibnamefont
  {Ben-David}}, \bibinfo {author} {\bibfnamefont {G.}~\bibnamefont {Cohen}},\
  and\ \bibinfo {author} {\bibfnamefont {J.}~\bibnamefont {Fineberg}},\
  }\bibfield  {title} {\bibinfo {title} {{The Dynamics of the Onset of
  Frictional Slip}},\ }\href {https://doi.org/10.1126/science.1194777}
  {\bibfield  {journal} {\bibinfo  {journal} {Science}\ }\textbf {\bibinfo
  {volume} {330}},\ \bibinfo {pages} {211} (\bibinfo {year}
  {2010})}\BibitemShut {NoStop}%
\bibitem [{\citenamefont {Kammer}\ \emph {et~al.}(2018)\citenamefont {Kammer},
  \citenamefont {Svetlizky}, \citenamefont {Cohen},\ and\ \citenamefont
  {Fineberg}}]{Kammer2018}%
  \BibitemOpen
  \bibfield  {author} {\bibinfo {author} {\bibfnamefont {D.~S.}\ \bibnamefont
  {Kammer}}, \bibinfo {author} {\bibfnamefont {I.}~\bibnamefont {Svetlizky}},
  \bibinfo {author} {\bibfnamefont {G.}~\bibnamefont {Cohen}},\ and\ \bibinfo
  {author} {\bibfnamefont {J.}~\bibnamefont {Fineberg}},\ }\bibfield  {title}
  {\bibinfo {title} {{The equation of motion for supershear frictional rupture
  fronts}},\ }\href {https://doi.org/10.1126/sciadv.aat5622} {\bibfield
  {journal} {\bibinfo  {journal} {Sci. Adv.}\ }\textbf {\bibinfo {volume}
  {4}},\ \bibinfo {pages} {eaat5622} (\bibinfo {year} {2018})}\BibitemShut
  {NoStop}%
\bibitem [{\citenamefont {Passel{\`e}gue}\ \emph {et~al.}(2013)\citenamefont
  {Passel{\`e}gue}, \citenamefont {Schubnel}, \citenamefont {Nielsen},
  \citenamefont {Bhat},\ and\ \citenamefont {Madariaga}}]{passelegue2013sub}%
  \BibitemOpen
  \bibfield  {author} {\bibinfo {author} {\bibfnamefont {F.~X.}\ \bibnamefont
  {Passel{\`e}gue}}, \bibinfo {author} {\bibfnamefont {A.}~\bibnamefont
  {Schubnel}}, \bibinfo {author} {\bibfnamefont {S.}~\bibnamefont {Nielsen}},
  \bibinfo {author} {\bibfnamefont {H.~S.}\ \bibnamefont {Bhat}},\ and\
  \bibinfo {author} {\bibfnamefont {R.}~\bibnamefont {Madariaga}},\ }\bibfield
  {title} {\bibinfo {title} {{From sub-Rayleigh to supershear ruptures during
  stick-slip experiments on crustal rocks}},\ }\href
  {https://www.science.org/doi/10.1126/science.1235637} {\bibfield  {journal}
  {\bibinfo  {journal} {Science}\ }\textbf {\bibinfo {volume} {340}},\ \bibinfo
  {pages} {1208} (\bibinfo {year} {2013})}\BibitemShut {NoStop}%
\bibitem [{\citenamefont {Xu}\ \emph {et~al.}(2018)\citenamefont {Xu},
  \citenamefont {Fukuyama}, \citenamefont {Yamashita}, \citenamefont
  {Mizoguchi}, \citenamefont {Takizawa},\ and\ \citenamefont
  {Kawakata}}]{xu2018strain}%
  \BibitemOpen
  \bibfield  {author} {\bibinfo {author} {\bibfnamefont {S.}~\bibnamefont
  {Xu}}, \bibinfo {author} {\bibfnamefont {E.}~\bibnamefont {Fukuyama}},
  \bibinfo {author} {\bibfnamefont {F.}~\bibnamefont {Yamashita}}, \bibinfo
  {author} {\bibfnamefont {K.}~\bibnamefont {Mizoguchi}}, \bibinfo {author}
  {\bibfnamefont {S.}~\bibnamefont {Takizawa}},\ and\ \bibinfo {author}
  {\bibfnamefont {H.}~\bibnamefont {Kawakata}},\ }\bibfield  {title} {\bibinfo
  {title} {Strain rate effect on fault slip and rupture evolution: Insight from
  meter-scale rock friction experiments},\ }\href
  {https://www.sciencedirect.com/science/article/pii/S0040195117305012}
  {\bibfield  {journal} {\bibinfo  {journal} {Tectonophysics}\ }\textbf
  {\bibinfo {volume} {733}},\ \bibinfo {pages} {209} (\bibinfo {year}
  {2018})}\BibitemShut {NoStop}%
\bibitem [{\citenamefont {Dieterich}(1979)}]{Dieterich1979}%
  \BibitemOpen
  \bibfield  {author} {\bibinfo {author} {\bibfnamefont {J.~H.}\ \bibnamefont
  {Dieterich}},\ }\bibfield  {title} {\bibinfo {title} {{Modeling of rock
  friction: 1. Experimental results and constitutive equations}},\ }\href
  {https://doi.org/10.1029/JB084iB05p02161} {\bibfield  {journal} {\bibinfo
  {journal} {J. Geophys. Res. Solid Earth}\ }\textbf {\bibinfo {volume} {84}},\
  \bibinfo {pages} {2161} (\bibinfo {year} {1979})}\BibitemShut {NoStop}%
\bibitem [{\citenamefont {Ruina}(1983)}]{Ruina1983}%
  \BibitemOpen
  \bibfield  {author} {\bibinfo {author} {\bibfnamefont {A.~L.}\ \bibnamefont
  {Ruina}},\ }\bibfield  {title} {\bibinfo {title} {Slip instability and state
  variable friction laws},\ }\href {https://doi.org/10.1029/JB088iB12p10359}
  {\bibfield  {journal} {\bibinfo  {journal} {J. Geophys. Res.}\ }\textbf
  {\bibinfo {volume} {88}},\ \bibinfo {pages} {10359} (\bibinfo {year}
  {1983})}\BibitemShut {NoStop}%
\bibitem [{\citenamefont {Shimamoto}(1986)}]{Shimamoto1986}%
  \BibitemOpen
  \bibfield  {author} {\bibinfo {author} {\bibfnamefont {T.}~\bibnamefont
  {Shimamoto}},\ }\bibfield  {title} {\bibinfo {title} {{Transition Between
  Frictional Slip and Ductile Flow for Halite Shear Zones at Room
  Temperature}},\ }\href {https://doi.org/10.1126/science.231.4739.711}
  {\bibfield  {journal} {\bibinfo  {journal} {Science}\ }\textbf {\bibinfo
  {volume} {231}},\ \bibinfo {pages} {711} (\bibinfo {year}
  {1986})}\BibitemShut {NoStop}%
\bibitem [{\citenamefont {Tullis}\ and\ \citenamefont
  {Weeks}(1986)}]{tullis1986constitutive}%
  \BibitemOpen
  \bibfield  {author} {\bibinfo {author} {\bibfnamefont {T.~E.}\ \bibnamefont
  {Tullis}}\ and\ \bibinfo {author} {\bibfnamefont {J.~D.}\ \bibnamefont
  {Weeks}},\ }\bibfield  {title} {\bibinfo {title} {Constitutive behavior and
  stability of frictional sliding of granite},\ }\href
  {https://link.springer.com/article/10.1007/BF00877209} {\bibfield  {journal}
  {\bibinfo  {journal} {Pure and Applied Geophysics}\ }\textbf {\bibinfo
  {volume} {124}},\ \bibinfo {pages} {383} (\bibinfo {year}
  {1986})}\BibitemShut {NoStop}%
\bibitem [{\citenamefont {Blanpied}\ \emph {et~al.}(1991)\citenamefont
  {Blanpied}, \citenamefont {Lockner},\ and\ \citenamefont
  {Byerlee}}]{blanpied1991fault}%
  \BibitemOpen
  \bibfield  {author} {\bibinfo {author} {\bibfnamefont {M.}~\bibnamefont
  {Blanpied}}, \bibinfo {author} {\bibfnamefont {D.}~\bibnamefont {Lockner}},\
  and\ \bibinfo {author} {\bibfnamefont {J.}~\bibnamefont {Byerlee}},\
  }\bibfield  {title} {\bibinfo {title} {Fault stability inferred from granite
  sliding experiments at hydrothermal conditions},\ }\href
  {https://doi.org/10.1029/91GL00469} {\bibfield  {journal} {\bibinfo
  {journal} {Geophysical Research Letters}\ }\textbf {\bibinfo {volume} {18}},\
  \bibinfo {pages} {609} (\bibinfo {year} {1991})}\BibitemShut {NoStop}%
\bibitem [{\citenamefont {Persson}(1998)}]{Persson1998}%
  \BibitemOpen
  \bibfield  {author} {\bibinfo {author} {\bibfnamefont {B.~N.~J.}\
  \bibnamefont {Persson}},\ }\href
  {https://link.springer.com/book/10.1007/978-3-662-04283-0} {\emph {\bibinfo
  {title} {{Sliding Friction: Physical Principles and Applications}}}}\
  (\bibinfo  {publisher} {Springer Sciense {\&} Buisness Media},\ \bibinfo
  {year} {1998})\BibitemShut {NoStop}%
\bibitem [{\citenamefont {Marone}(1998)}]{Marone1998a}%
  \BibitemOpen
  \bibfield  {author} {\bibinfo {author} {\bibfnamefont {C.}~\bibnamefont
  {Marone}},\ }\bibfield  {title} {\bibinfo {title} {{Laboratoty-derived
  friction laws and their application to seismic faulting}},\ }\href
  {https://doi.org/10.1146/annurev.earth.26.1.643} {\bibfield  {journal}
  {\bibinfo  {journal} {Annu. Rev. Earth Planet. Sci.}\ }\textbf {\bibinfo
  {volume} {26}},\ \bibinfo {pages} {643} (\bibinfo {year} {1998})}\BibitemShut
  {NoStop}%
\bibitem [{\citenamefont {Baumberger}\ and\ \citenamefont
  {Berthoud}(1999)}]{Baumberger1999}%
  \BibitemOpen
  \bibfield  {author} {\bibinfo {author} {\bibfnamefont {T.}~\bibnamefont
  {Baumberger}}\ and\ \bibinfo {author} {\bibfnamefont {P.}~\bibnamefont
  {Berthoud}},\ }\bibfield  {title} {\bibinfo {title} {{Physical analysis of
  the state- and rate-dependent friction law. II. Dynamic friction}},\ }\href
  {https://doi.org/10.1103/PhysRevB.60.3928} {\bibfield  {journal} {\bibinfo
  {journal} {Phys. Rev. B}\ }\textbf {\bibinfo {volume} {60}},\ \bibinfo
  {pages} {3928} (\bibinfo {year} {1999})}\BibitemShut {NoStop}%
\bibitem [{\citenamefont {Nakatani}(2001)}]{Nakatani2001}%
  \BibitemOpen
  \bibfield  {author} {\bibinfo {author} {\bibfnamefont {M.}~\bibnamefont
  {Nakatani}},\ }\bibfield  {title} {\bibinfo {title} {{Conceptual and physical
  clarification of rate and state friction: Frictional sliding as a thermally
  activated rheology}},\ }\href {https://doi.org/10.1029/2000JB900453}
  {\bibfield  {journal} {\bibinfo  {journal} {J. Geophys. Res. Solid Earth}\
  }\textbf {\bibinfo {volume} {106}},\ \bibinfo {pages} {13347} (\bibinfo
  {year} {2001})}\BibitemShut {NoStop}%
\bibitem [{\citenamefont {Baumberger}\ and\ \citenamefont
  {Caroli}(2006)}]{Baumberger2006}%
  \BibitemOpen
  \bibfield  {author} {\bibinfo {author} {\bibfnamefont {T.}~\bibnamefont
  {Baumberger}}\ and\ \bibinfo {author} {\bibfnamefont {C.}~\bibnamefont
  {Caroli}},\ }\bibfield  {title} {\bibinfo {title} {{Solid friction from
  stick-slip down to pinning and aging}},\ }\href
  {https://doi.org/10.1080/00018730600732186} {\bibfield  {journal} {\bibinfo
  {journal} {Adv. Phys.}\ }\textbf {\bibinfo {volume} {55}},\ \bibinfo {pages}
  {279} (\bibinfo {year} {2006})}\BibitemShut {NoStop}%
\bibitem [{\citenamefont {Reches}\ and\ \citenamefont
  {Lockner}(2010)}]{Reches2010}%
  \BibitemOpen
  \bibfield  {author} {\bibinfo {author} {\bibfnamefont {Z.}~\bibnamefont
  {Reches}}\ and\ \bibinfo {author} {\bibfnamefont {D.~A.}\ \bibnamefont
  {Lockner}},\ }\bibfield  {title} {\bibinfo {title} {{Fault weakening and
  earthquake instability by powder lubrication}},\ }\href
  {https://doi.org/10.1038/nature09348} {\bibfield  {journal} {\bibinfo
  {journal} {Nature}\ }\textbf {\bibinfo {volume} {467}},\ \bibinfo {pages}
  {452} (\bibinfo {year} {2010})}\BibitemShut {NoStop}%
\bibitem [{\citenamefont {Bar-Sinai}\ \emph {et~al.}(2014)\citenamefont
  {Bar-Sinai}, \citenamefont {Spatschek}, \citenamefont {Brener},\ and\
  \citenamefont {Bouchbinder}}]{Bar-Sinai2014}%
  \BibitemOpen
  \bibfield  {author} {\bibinfo {author} {\bibfnamefont {Y.}~\bibnamefont
  {Bar-Sinai}}, \bibinfo {author} {\bibfnamefont {R.}~\bibnamefont
  {Spatschek}}, \bibinfo {author} {\bibfnamefont {E.~A.}\ \bibnamefont
  {Brener}},\ and\ \bibinfo {author} {\bibfnamefont {E.}~\bibnamefont
  {Bouchbinder}},\ }\bibfield  {title} {\bibinfo {title} {{On the
  velocity-strengthening behavior of dry friction}},\ }\href
  {https://doi.org/10.1002/2013JB010586} {\bibfield  {journal} {\bibinfo
  {journal} {J. Geophys. Res. Solid Earth}\ }\textbf {\bibinfo {volume}
  {119}},\ \bibinfo {pages} {1738} (\bibinfo {year} {2014})}\BibitemShut
  {NoStop}%
\bibitem [{\citenamefont {Gou}\ \emph {et~al.}(2024)\citenamefont {Gou},
  \citenamefont {Hu}, \citenamefont {Xu}, \citenamefont {Chen}, \citenamefont
  {McSaveney}, \citenamefont {Breard}, \citenamefont {Huang}, \citenamefont
  {Wang}, \citenamefont {Jia},\ and\ \citenamefont {Zhou}}]{gou2024variation}%
  \BibitemOpen
  \bibfield  {author} {\bibinfo {author} {\bibfnamefont {H.}~\bibnamefont
  {Gou}}, \bibinfo {author} {\bibfnamefont {W.}~\bibnamefont {Hu}}, \bibinfo
  {author} {\bibfnamefont {Q.}~\bibnamefont {Xu}}, \bibinfo {author}
  {\bibfnamefont {J.}~\bibnamefont {Chen}}, \bibinfo {author} {\bibfnamefont
  {M.}~\bibnamefont {McSaveney}}, \bibinfo {author} {\bibfnamefont {E.~C.}\
  \bibnamefont {Breard}}, \bibinfo {author} {\bibfnamefont {R.}~\bibnamefont
  {Huang}}, \bibinfo {author} {\bibfnamefont {Y.}~\bibnamefont {Wang}},
  \bibinfo {author} {\bibfnamefont {X.}~\bibnamefont {Jia}},\ and\ \bibinfo
  {author} {\bibfnamefont {L.}~\bibnamefont {Zhou}},\ }\bibfield  {title}
  {\bibinfo {title} {Variation in granular frictional resistance across nine
  orders of magnitude in shear velocity},\ }\href
  {https://doi.org/10.1029/2023JB028241} {\bibfield  {journal} {\bibinfo
  {journal} {Journal of Geophysical Research: Solid Earth}\ }\textbf {\bibinfo
  {volume} {129}},\ \bibinfo {pages} {e2023JB028241} (\bibinfo {year}
  {2024})}\BibitemShut {NoStop}%
\bibitem [{\citenamefont {Viesca}\ and\ \citenamefont
  {Garagash}(2015)}]{Viesca2015}%
  \BibitemOpen
  \bibfield  {author} {\bibinfo {author} {\bibfnamefont {R.~C.}\ \bibnamefont
  {Viesca}}\ and\ \bibinfo {author} {\bibfnamefont {D.~I.}\ \bibnamefont
  {Garagash}},\ }\bibfield  {title} {\bibinfo {title} {{Ubiquitous weakening of
  faults due to thermal pressurization}},\ }\href
  {https://doi.org/10.1038/ngeo2554} {\bibfield  {journal} {\bibinfo  {journal}
  {Nat. Geosci.}\ }\textbf {\bibinfo {volume} {8}},\ \bibinfo {pages} {875}
  (\bibinfo {year} {2015})}\BibitemShut {NoStop}%
\bibitem [{\citenamefont {Brantut}\ and\ \citenamefont
  {Viesca}(2017)}]{brantut2017fracture}%
  \BibitemOpen
  \bibfield  {author} {\bibinfo {author} {\bibfnamefont {N.}~\bibnamefont
  {Brantut}}\ and\ \bibinfo {author} {\bibfnamefont {R.~C.}\ \bibnamefont
  {Viesca}},\ }\bibfield  {title} {\bibinfo {title} {The fracture energy of
  ruptures driven by flash heating},\ }\href
  {https://doi.org/10.1002/2017GL074110} {\bibfield  {journal} {\bibinfo
  {journal} {Geophysical Research Letters}\ }\textbf {\bibinfo {volume} {44}},\
  \bibinfo {pages} {6718} (\bibinfo {year} {2017})}\BibitemShut {NoStop}%
\bibitem [{\citenamefont {Bar-Sinai}\ \emph {et~al.}(2019)\citenamefont
  {Bar-Sinai}, \citenamefont {Aldam}, \citenamefont {Spatschek}, \citenamefont
  {Brener},\ and\ \citenamefont {Bouchbinder}}]{bar2019spatiotemporal}%
  \BibitemOpen
  \bibfield  {author} {\bibinfo {author} {\bibfnamefont {Y.}~\bibnamefont
  {Bar-Sinai}}, \bibinfo {author} {\bibfnamefont {M.}~\bibnamefont {Aldam}},
  \bibinfo {author} {\bibfnamefont {R.}~\bibnamefont {Spatschek}}, \bibinfo
  {author} {\bibfnamefont {E.~A.}\ \bibnamefont {Brener}},\ and\ \bibinfo
  {author} {\bibfnamefont {E.}~\bibnamefont {Bouchbinder}},\ }\bibfield
  {title} {\bibinfo {title} {{Spatiotemporal dynamics of frictional systems:
  The interplay of interfacial friction and bulk elasticity}},\ }\href
  {https://www.mdpi.com/2075-4442/7/10/91} {\bibfield  {journal} {\bibinfo
  {journal} {Lubricants}\ }\textbf {\bibinfo {volume} {7}},\ \bibinfo {pages}
  {91} (\bibinfo {year} {2019})}\BibitemShut {NoStop}%
\bibitem [{\citenamefont {Landau}\ and\ \citenamefont
  {Lifshitz}(1986)}]{Landau1986}%
  \BibitemOpen
  \bibfield  {author} {\bibinfo {author} {\bibfnamefont {L.}~\bibnamefont
  {Landau}}\ and\ \bibinfo {author} {\bibfnamefont {E.}~\bibnamefont
  {Lifshitz}},\ }\href@noop {} {\emph {\bibinfo {title} {{Theory of Elasticity,
  Third Edition: Volume 7 (Course of Theoretical Physics)}}}}\ (\bibinfo
  {publisher} {Butterworth-Heinemann},\ \bibinfo {year} {1986})\BibitemShut
  {NoStop}%
\bibitem [{Note1()}]{Note1}%
  \BibitemOpen
  \bibinfo {note} {See Sect.~4.3 in~\cite {Freund1998}, Eq.~(4.3.2) therein in
  particular.}\BibitemShut {Stop}%
\bibitem [{Note2()}]{Note2}%
  \BibitemOpen
  \bibinfo {note} {In Eq.~(4.3.2) of~\cite {Freund1998}, displacement
  potentials are used such that the order of singularity therein is
  $p_{_{0}}\protect \!=\protect \!\xi +2$, with $1\protect \!<\protect
  \!p_{_{0}}\protect \!<\protect \!2$.}\BibitemShut {Stop}%
\bibitem [{\citenamefont {Palmer}\ and\ \citenamefont
  {Rice}(1973)}]{Palmer1973}%
  \BibitemOpen
  \bibfield  {author} {\bibinfo {author} {\bibfnamefont {A.~C.}\ \bibnamefont
  {Palmer}}\ and\ \bibinfo {author} {\bibfnamefont {J.~R.}\ \bibnamefont
  {Rice}},\ }\bibfield  {title} {\bibinfo {title} {{The growth of slip surfaces
  in the progressive failure of over-consolidated clay}},\ }\href
  {https://doi.org/10.1098/rspa.1973.0040} {\bibfield  {journal} {\bibinfo
  {journal} {Proc. R. Soc. A Math. Phys. Eng. Sci.}\ }\textbf {\bibinfo
  {volume} {332}},\ \bibinfo {pages} {527} (\bibinfo {year}
  {1973})}\BibitemShut {NoStop}%
\bibitem [{\citenamefont {Berman}\ \emph {et~al.}(2020)\citenamefont {Berman},
  \citenamefont {Cohen},\ and\ \citenamefont {Fineberg}}]{berman2020dynamics}%
  \BibitemOpen
  \bibfield  {author} {\bibinfo {author} {\bibfnamefont {N.}~\bibnamefont
  {Berman}}, \bibinfo {author} {\bibfnamefont {G.}~\bibnamefont {Cohen}},\ and\
  \bibinfo {author} {\bibfnamefont {J.}~\bibnamefont {Fineberg}},\ }\bibfield
  {title} {\bibinfo {title} {Dynamics and properties of the cohesive zone in
  rapid fracture and friction},\ }\href
  {https://journals.aps.org/prl/abstract/10.1103/PhysRevLett.125.125503}
  {\bibfield  {journal} {\bibinfo  {journal} {Physical Review Letters}\
  }\textbf {\bibinfo {volume} {125}},\ \bibinfo {pages} {125503} (\bibinfo
  {year} {2020})}\BibitemShut {NoStop}%
\bibitem [{\citenamefont {Kanamori}\ and\ \citenamefont
  {Heaton}(2000)}]{kanamori2000microscopic}%
  \BibitemOpen
  \bibfield  {author} {\bibinfo {author} {\bibfnamefont {H.}~\bibnamefont
  {Kanamori}}\ and\ \bibinfo {author} {\bibfnamefont {T.~H.}\ \bibnamefont
  {Heaton}},\ }\bibfield  {title} {\bibinfo {title} {Microscopic and
  macroscopic physics of earthquakes},\ }\href
  {https://agupubs.onlinelibrary.wiley.com/doi/book/10.1029/GM120} {\bibfield
  {journal} {\bibinfo  {journal} {Geocomplexity and the Physics of
  Earthquakes}\ }\textbf {\bibinfo {volume} {120}},\ \bibinfo {pages} {147}
  (\bibinfo {year} {2000})}\BibitemShut {NoStop}%
\bibitem [{\citenamefont {Abercrombie}\ and\ \citenamefont
  {Rice}(2005)}]{Abercrombie2005}%
  \BibitemOpen
  \bibfield  {author} {\bibinfo {author} {\bibfnamefont {R.~E.}\ \bibnamefont
  {Abercrombie}}\ and\ \bibinfo {author} {\bibfnamefont {J.~R.}\ \bibnamefont
  {Rice}},\ }\bibfield  {title} {\bibinfo {title} {{Can observations of
  earthquake scaling constrain slip weakening?}},\ }\href
  {https://doi.org/10.1111/j.1365-246X.2005.02579.x} {\bibfield  {journal}
  {\bibinfo  {journal} {Geophys. J. Int.}\ }\textbf {\bibinfo {volume} {162}},\
  \bibinfo {pages} {406} (\bibinfo {year} {2005})}\BibitemShut {NoStop}%
\bibitem [{\citenamefont {Tinti}\ \emph {et~al.}(2005)\citenamefont {Tinti},
  \citenamefont {Spudich},\ and\ \citenamefont {Cocco}}]{Tinti2005}%
  \BibitemOpen
  \bibfield  {author} {\bibinfo {author} {\bibfnamefont {E.}~\bibnamefont
  {Tinti}}, \bibinfo {author} {\bibfnamefont {P.}~\bibnamefont {Spudich}},\
  and\ \bibinfo {author} {\bibfnamefont {M.}~\bibnamefont {Cocco}},\ }\bibfield
   {title} {\bibinfo {title} {{Earthquake fracture energy inferred from
  kinematic rupture models on extended faults}},\ }\href
  {https://doi.org/10.1029/2005JB003644} {\bibfield  {journal} {\bibinfo
  {journal} {J. Geophys. Res.}\ }\textbf {\bibinfo {volume} {110}},\ \bibinfo
  {pages} {B12303} (\bibinfo {year} {2005})}\BibitemShut {NoStop}%
\bibitem [{\citenamefont {Nielsen}\ \emph {et~al.}(2016)\citenamefont
  {Nielsen}, \citenamefont {Spagnuolo}, \citenamefont {Smith}, \citenamefont
  {Violay}, \citenamefont {{Di Toro}},\ and\ \citenamefont
  {Bistacchi}}]{Nielsen2016}%
  \BibitemOpen
  \bibfield  {author} {\bibinfo {author} {\bibfnamefont {S.~B.}\ \bibnamefont
  {Nielsen}}, \bibinfo {author} {\bibfnamefont {E.}~\bibnamefont {Spagnuolo}},
  \bibinfo {author} {\bibfnamefont {S.~A.~F.}\ \bibnamefont {Smith}}, \bibinfo
  {author} {\bibfnamefont {M.}~\bibnamefont {Violay}}, \bibinfo {author}
  {\bibfnamefont {G.}~\bibnamefont {{Di Toro}}},\ and\ \bibinfo {author}
  {\bibfnamefont {A.}~\bibnamefont {Bistacchi}},\ }\bibfield  {title} {\bibinfo
  {title} {{Scaling in natural and laboratory earthquakes}},\ }\href
  {https://doi.org/10.1002/2015GL067490} {\bibfield  {journal} {\bibinfo
  {journal} {Geophys. Res. Lett.}\ }\textbf {\bibinfo {volume} {43}},\ \bibinfo
  {pages} {1504} (\bibinfo {year} {2016})}\BibitemShut {NoStop}%
\bibitem [{\citenamefont {Brener}\ and\ \citenamefont
  {Bouchbinder}(2021{\natexlab{a}})}]{Brener2021unconventional}%
  \BibitemOpen
  \bibfield  {author} {\bibinfo {author} {\bibfnamefont {E.~A.}\ \bibnamefont
  {Brener}}\ and\ \bibinfo {author} {\bibfnamefont {E.}~\bibnamefont
  {Bouchbinder}},\ }\bibfield  {title} {\bibinfo {title} {Unconventional
  singularities and energy balance in frictional rupture},\ }\href
  {https://www.nature.com/articles/s41467-021-22806-9} {\bibfield  {journal}
  {\bibinfo  {journal} {Nat. Commun.}\ }\textbf {\bibinfo {volume} {12}},\
  \bibinfo {pages} {1} (\bibinfo {year} {2021}{\natexlab{a}})}\BibitemShut
  {NoStop}%
\bibitem [{\citenamefont {Brener}\ and\ \citenamefont
  {Bouchbinder}(2021{\natexlab{b}})}]{Brener2021JMPS}%
  \BibitemOpen
  \bibfield  {author} {\bibinfo {author} {\bibfnamefont {E.~A.}\ \bibnamefont
  {Brener}}\ and\ \bibinfo {author} {\bibfnamefont {E.}~\bibnamefont
  {Bouchbinder}},\ }\bibfield  {title} {\bibinfo {title} {Theory of
  unconventional singularities of frictional shear cracks},\ }\href
  {https://doi.org/https://doi.org/10.1016/j.jmps.2021.104466} {\bibfield
  {journal} {\bibinfo  {journal} {J. Mech. Phys. Solids}\ }\textbf {\bibinfo
  {volume} {153}},\ \bibinfo {pages} {104466} (\bibinfo {year}
  {2021}{\natexlab{b}})}\BibitemShut {NoStop}%
\bibitem [{\citenamefont {Cocco}\ \emph {et~al.}(2023)\citenamefont {Cocco},
  \citenamefont {Aretusini}, \citenamefont {Cornelio}, \citenamefont {Nielsen},
  \citenamefont {Spagnuolo}, \citenamefont {Tinti},\ and\ \citenamefont
  {Di~Toro}}]{cocco2023fracture}%
  \BibitemOpen
  \bibfield  {author} {\bibinfo {author} {\bibfnamefont {M.}~\bibnamefont
  {Cocco}}, \bibinfo {author} {\bibfnamefont {S.}~\bibnamefont {Aretusini}},
  \bibinfo {author} {\bibfnamefont {C.}~\bibnamefont {Cornelio}}, \bibinfo
  {author} {\bibfnamefont {S.~B.}\ \bibnamefont {Nielsen}}, \bibinfo {author}
  {\bibfnamefont {E.}~\bibnamefont {Spagnuolo}}, \bibinfo {author}
  {\bibfnamefont {E.}~\bibnamefont {Tinti}},\ and\ \bibinfo {author}
  {\bibfnamefont {G.}~\bibnamefont {Di~Toro}},\ }\bibfield  {title} {\bibinfo
  {title} {Fracture energy and breakdown work during earthquakes},\ }\href
  {https://www.annualreviews.org/content/journals/10.1146/annurev-earth-071822-100304}
  {\bibfield  {journal} {\bibinfo  {journal} {Annual Review of Earth and
  Planetary Sciences}\ }\textbf {\bibinfo {volume} {51}},\ \bibinfo {pages}
  {217} (\bibinfo {year} {2023})}\BibitemShut {NoStop}%
\bibitem [{\citenamefont {Ida}(1974)}]{Ida1974slow}%
  \BibitemOpen
  \bibfield  {author} {\bibinfo {author} {\bibfnamefont {Y.}~\bibnamefont
  {Ida}},\ }\bibfield  {title} {\bibinfo {title} {Slow-moving deformation
  pulses along tectonic faults},\ }\href
  {https://www.sciencedirect.com/science/article/pii/0031920174900600}
  {\bibfield  {journal} {\bibinfo  {journal} {Physics of the Earth and
  Planetary Interiors}\ }\textbf {\bibinfo {volume} {9}},\ \bibinfo {pages}
  {328} (\bibinfo {year} {1974})}\BibitemShut {NoStop}%
\bibitem [{\citenamefont {Brener}\ and\ \citenamefont
  {Marchenko}(2002)}]{Brener2002}%
  \BibitemOpen
  \bibfield  {author} {\bibinfo {author} {\bibfnamefont {E.~A.}\ \bibnamefont
  {Brener}}\ and\ \bibinfo {author} {\bibfnamefont {V.~I.}\ \bibnamefont
  {Marchenko}},\ }\bibfield  {title} {\bibinfo {title} {{Frictional shear
  cracks}},\ }\href {https://doi.org/10.1134/1.1517386} {\bibfield  {journal}
  {\bibinfo  {journal} {J. Exp. Theor. Phys. Lett.}\ }\textbf {\bibinfo
  {volume} {76}},\ \bibinfo {pages} {211} (\bibinfo {year} {2002})}\BibitemShut
  {NoStop}%
\bibitem [{\citenamefont {Brener}\ \emph {et~al.}(2005)\citenamefont {Brener},
  \citenamefont {Malinin},\ and\ \citenamefont {Marchenko}}]{Brener2005}%
  \BibitemOpen
  \bibfield  {author} {\bibinfo {author} {\bibfnamefont {E.~A.}\ \bibnamefont
  {Brener}}, \bibinfo {author} {\bibfnamefont {S.~V.}\ \bibnamefont
  {Malinin}},\ and\ \bibinfo {author} {\bibfnamefont {V.~I.}\ \bibnamefont
  {Marchenko}},\ }\bibfield  {title} {\bibinfo {title} {{Fracture and friction:
  Stick-slip motion}},\ }\href {https://doi.org/10.1140/epje/i2004-10112-3}
  {\bibfield  {journal} {\bibinfo  {journal} {Eur. Phys. J. E}\ }\textbf
  {\bibinfo {volume} {17}},\ \bibinfo {pages} {101} (\bibinfo {year}
  {2005})}\BibitemShut {NoStop}%
\bibitem [{\citenamefont {Desroches}\ \emph {et~al.}(1994)\citenamefont
  {Desroches}, \citenamefont {Detournay}, \citenamefont {Lenoach},
  \citenamefont {Papanastasiou}, \citenamefont {Pearson}, \citenamefont
  {Thiercelin},\ and\ \citenamefont {Cheng}}]{Desroches1994}%
  \BibitemOpen
  \bibfield  {author} {\bibinfo {author} {\bibfnamefont {J.}~\bibnamefont
  {Desroches}}, \bibinfo {author} {\bibfnamefont {E.}~\bibnamefont
  {Detournay}}, \bibinfo {author} {\bibfnamefont {B.}~\bibnamefont {Lenoach}},
  \bibinfo {author} {\bibfnamefont {P.}~\bibnamefont {Papanastasiou}}, \bibinfo
  {author} {\bibfnamefont {J.~R.~A.}\ \bibnamefont {Pearson}}, \bibinfo
  {author} {\bibfnamefont {M.}~\bibnamefont {Thiercelin}},\ and\ \bibinfo
  {author} {\bibfnamefont {A.}~\bibnamefont {Cheng}},\ }\bibfield  {title}
  {\bibinfo {title} {The crack tip region in hydraulic fracturing},\ }\href
  {https://doi.org/10.1098/rspa.1994.0127} {\bibfield  {journal} {\bibinfo
  {journal} {Proceedings of the Royal Society of London. Series A: Mathematical
  and Physical Sciences}\ }\textbf {\bibinfo {volume} {447}},\ \bibinfo {pages}
  {39} (\bibinfo {year} {1994})}\BibitemShut {NoStop}%
\bibitem [{\citenamefont {Garagash}\ and\ \citenamefont
  {Detournay}(2000)}]{garagash2000tip}%
  \BibitemOpen
  \bibfield  {author} {\bibinfo {author} {\bibfnamefont {D.}~\bibnamefont
  {Garagash}}\ and\ \bibinfo {author} {\bibfnamefont {E.}~\bibnamefont
  {Detournay}},\ }\bibfield  {title} {\bibinfo {title} {The tip region of a
  fluid-driven fracture in an elastic medium},\ }\href
  {https://doi.org/10.1115/1.321162} {\bibfield  {journal} {\bibinfo  {journal}
  {J. Appl. Mech.}\ }\textbf {\bibinfo {volume} {67}},\ \bibinfo {pages} {183}
  (\bibinfo {year} {2000})}\BibitemShut {NoStop}%
\bibitem [{\citenamefont {Detournay}(2016)}]{detournay2016mechanics}%
  \BibitemOpen
  \bibfield  {author} {\bibinfo {author} {\bibfnamefont {E.}~\bibnamefont
  {Detournay}},\ }\bibfield  {title} {\bibinfo {title} {Mechanics of hydraulic
  fractures},\ }\href {https://doi.org/10.1146/annurev-fluid-010814-014736}
  {\bibfield  {journal} {\bibinfo  {journal} {Annual review of fluid
  mechanics}\ }\textbf {\bibinfo {volume} {48}},\ \bibinfo {pages} {311}
  (\bibinfo {year} {2016})}\BibitemShut {NoStop}%
\bibitem [{\citenamefont {Barras}\ \emph {et~al.}(2020)\citenamefont {Barras},
  \citenamefont {Aldam}, \citenamefont {Roch}, \citenamefont {Brener},
  \citenamefont {Bouchbinder},\ and\ \citenamefont {Molinari}}]{Barras2020}%
  \BibitemOpen
  \bibfield  {author} {\bibinfo {author} {\bibfnamefont {F.}~\bibnamefont
  {Barras}}, \bibinfo {author} {\bibfnamefont {M.}~\bibnamefont {Aldam}},
  \bibinfo {author} {\bibfnamefont {T.}~\bibnamefont {Roch}}, \bibinfo {author}
  {\bibfnamefont {E.~A.}\ \bibnamefont {Brener}}, \bibinfo {author}
  {\bibfnamefont {E.}~\bibnamefont {Bouchbinder}},\ and\ \bibinfo {author}
  {\bibfnamefont {J.-F.}\ \bibnamefont {Molinari}},\ }\bibfield  {title}
  {\bibinfo {title} {{The emergence of crack-like behavior of frictional
  rupture: Edge singularity and energy balance}},\ }\href
  {https://www.sciencedirect.com/science/article/pii/S0012821X19306703}
  {\bibfield  {journal} {\bibinfo  {journal} {Earth and Planetary Science
  Letters}\ }\textbf {\bibinfo {volume} {531}},\ \bibinfo {pages} {115978}
  (\bibinfo {year} {2020})}\BibitemShut {NoStop}%
\bibitem [{\citenamefont {Paglialunga}\ \emph {et~al.}(2022)\citenamefont
  {Paglialunga}, \citenamefont {Passel{\`e}gue}, \citenamefont {Brantut},
  \citenamefont {Barras}, \citenamefont {Lebihain},\ and\ \citenamefont
  {Violay}}]{paglialunga2022scale}%
  \BibitemOpen
  \bibfield  {author} {\bibinfo {author} {\bibfnamefont {F.}~\bibnamefont
  {Paglialunga}}, \bibinfo {author} {\bibfnamefont {F.~X.}\ \bibnamefont
  {Passel{\`e}gue}}, \bibinfo {author} {\bibfnamefont {N.}~\bibnamefont
  {Brantut}}, \bibinfo {author} {\bibfnamefont {F.}~\bibnamefont {Barras}},
  \bibinfo {author} {\bibfnamefont {M.}~\bibnamefont {Lebihain}},\ and\
  \bibinfo {author} {\bibfnamefont {M.}~\bibnamefont {Violay}},\ }\bibfield
  {title} {\bibinfo {title} {On the scale dependence in the dynamics of
  frictional rupture: Constant fracture energy versus size-dependent breakdown
  work},\ }\href
  {https://www.sciencedirect.com/science/article/pii/S0012821X22000784}
  {\bibfield  {journal} {\bibinfo  {journal} {Earth and Planetary Science
  Letters}\ }\textbf {\bibinfo {volume} {584}},\ \bibinfo {pages} {117442}
  (\bibinfo {year} {2022})}\BibitemShut {NoStop}%
\bibitem [{\citenamefont {Paglialunga}\ \emph {et~al.}(2024)\citenamefont
  {Paglialunga}, \citenamefont {Passel{\`e}gue}, \citenamefont {Lebihain},\
  and\ \citenamefont {Violay}}]{paglialunga2024frictional}%
  \BibitemOpen
  \bibfield  {author} {\bibinfo {author} {\bibfnamefont {F.}~\bibnamefont
  {Paglialunga}}, \bibinfo {author} {\bibfnamefont {F.}~\bibnamefont
  {Passel{\`e}gue}}, \bibinfo {author} {\bibfnamefont {M.}~\bibnamefont
  {Lebihain}},\ and\ \bibinfo {author} {\bibfnamefont {M.}~\bibnamefont
  {Violay}},\ }\bibfield  {title} {\bibinfo {title} {Frictional weakening leads
  to unconventional singularities during dynamic rupture propagation},\ }\href
  {https://www.sciencedirect.com/science/article/pii/S0012821X23005617}
  {\bibfield  {journal} {\bibinfo  {journal} {Earth and Planetary Science
  Letters}\ }\textbf {\bibinfo {volume} {626}},\ \bibinfo {pages} {118550}
  (\bibinfo {year} {2024})}\BibitemShut {NoStop}%
\bibitem [{\citenamefont {Fryer}\ \emph {et~al.}(2024)\citenamefont {Fryer},
  \citenamefont {Lebihain}, \citenamefont {No{\"e}l}, \citenamefont
  {Paglialunga},\ and\ \citenamefont {Passel{\`e}gue}}]{fryer2024effect}%
  \BibitemOpen
  \bibfield  {author} {\bibinfo {author} {\bibfnamefont {B.}~\bibnamefont
  {Fryer}}, \bibinfo {author} {\bibfnamefont {M.}~\bibnamefont {Lebihain}},
  \bibinfo {author} {\bibfnamefont {C.}~\bibnamefont {No{\"e}l}}, \bibinfo
  {author} {\bibfnamefont {F.}~\bibnamefont {Paglialunga}},\ and\ \bibinfo
  {author} {\bibfnamefont {F.}~\bibnamefont {Passel{\`e}gue}},\ }\bibfield
  {title} {\bibinfo {title} {The effect of stress barriers on
  unconventional-singularity-driven frictional rupture},\ }\href
  {https://www.sciencedirect.com/science/article/pii/S0022509624003429}
  {\bibfield  {journal} {\bibinfo  {journal} {Journal of the Mechanics and
  Physics of Solids}\ }\textbf {\bibinfo {volume} {193}},\ \bibinfo {pages}
  {105876} (\bibinfo {year} {2024})}\BibitemShut {NoStop}%
\bibitem [{\citenamefont {Roch}\ \emph {et~al.}(2022)\citenamefont {Roch},
  \citenamefont {Barras}, \citenamefont {Geubelle},\ and\ \citenamefont
  {Molinari}}]{roch2022cracklet}%
  \BibitemOpen
  \bibfield  {author} {\bibinfo {author} {\bibfnamefont {T.}~\bibnamefont
  {Roch}}, \bibinfo {author} {\bibfnamefont {F.}~\bibnamefont {Barras}},
  \bibinfo {author} {\bibfnamefont {P.~H.}\ \bibnamefont {Geubelle}},\ and\
  \bibinfo {author} {\bibfnamefont {J.-F.}\ \bibnamefont {Molinari}},\
  }\bibfield  {title} {\bibinfo {title} {{cRacklet: a spectral boundary
  integral method library for interfacial rupture simulation}},\ }\href
  {https://doi.org/10.21105/joss.03724} {\bibfield  {journal} {\bibinfo
  {journal} {Journal of Open Source Software}\ }\textbf {\bibinfo {volume}
  {7}},\ \bibinfo {pages} {3724} (\bibinfo {year} {2022})}\BibitemShut
  {NoStop}%
\bibitem [{\citenamefont {Festa}\ and\ \citenamefont
  {Vilotte}(2006)}]{festa2006influence}%
  \BibitemOpen
  \bibfield  {author} {\bibinfo {author} {\bibfnamefont {G.}~\bibnamefont
  {Festa}}\ and\ \bibinfo {author} {\bibfnamefont {J.-P.}\ \bibnamefont
  {Vilotte}},\ }\bibfield  {title} {\bibinfo {title} {{Influence of the rupture
  initiation on the intersonic transition: Crack-like versus pulse-like
  modes}},\ }\href
  {https://agupubs.onlinelibrary.wiley.com/doi/full/10.1029/2006GL026378}
  {\bibfield  {journal} {\bibinfo  {journal} {Geophysical Research Letters}\
  }\textbf {\bibinfo {volume} {33}},\ \bibinfo {pages} {L15320} (\bibinfo
  {year} {2006})}\BibitemShut {NoStop}%
\bibitem [{\citenamefont {Lu}\ \emph {et~al.}(2009)\citenamefont {Lu},
  \citenamefont {Lapusta},\ and\ \citenamefont {Rosakis}}]{lu2009analysis}%
  \BibitemOpen
  \bibfield  {author} {\bibinfo {author} {\bibfnamefont {X.}~\bibnamefont
  {Lu}}, \bibinfo {author} {\bibfnamefont {N.}~\bibnamefont {Lapusta}},\ and\
  \bibinfo {author} {\bibfnamefont {A.~J.}\ \bibnamefont {Rosakis}},\
  }\bibfield  {title} {\bibinfo {title} {Analysis of supershear transition
  regimes in rupture experiments: the effect of nucleation conditions and
  friction parameters},\ }\href
  {https://academic.oup.com/gji/article/177/2/717/2023688} {\bibfield
  {journal} {\bibinfo  {journal} {Geophysical Journal International}\ }\textbf
  {\bibinfo {volume} {177}},\ \bibinfo {pages} {717} (\bibinfo {year}
  {2009})}\BibitemShut {NoStop}%
\bibitem [{\citenamefont {Liu}\ \emph {et~al.}(2014)\citenamefont {Liu},
  \citenamefont {Bizzarri},\ and\ \citenamefont {Das}}]{liu2014progression}%
  \BibitemOpen
  \bibfield  {author} {\bibinfo {author} {\bibfnamefont {C.}~\bibnamefont
  {Liu}}, \bibinfo {author} {\bibfnamefont {A.}~\bibnamefont {Bizzarri}},\ and\
  \bibinfo {author} {\bibfnamefont {S.}~\bibnamefont {Das}},\ }\bibfield
  {title} {\bibinfo {title} {Progression of spontaneous in-plane shear faults
  from sub-rayleigh to compressional wave rupture speeds},\ }\href
  {https://doi.org/10.1002/2014JB011187} {\bibfield  {journal} {\bibinfo
  {journal} {Journal of Geophysical Research: Solid Earth}\ }\textbf {\bibinfo
  {volume} {119}},\ \bibinfo {pages} {8331} (\bibinfo {year}
  {2014})}\BibitemShut {NoStop}%
\bibitem [{\citenamefont {Bizzarri}\ and\ \citenamefont
  {Liu}(2016)}]{bizzarri2016near}%
  \BibitemOpen
  \bibfield  {author} {\bibinfo {author} {\bibfnamefont {A.}~\bibnamefont
  {Bizzarri}}\ and\ \bibinfo {author} {\bibfnamefont {C.}~\bibnamefont {Liu}},\
  }\bibfield  {title} {\bibinfo {title} {Near-field radiated wave field may
  help to understand the style of the supershear transition of dynamic
  ruptures},\ }\href
  {https://www.sciencedirect.com/science/article/pii/S0031920116300838?via%3Dihub}
  {\bibfield  {journal} {\bibinfo  {journal} {Physics of the Earth and
  Planetary Interiors}\ }\textbf {\bibinfo {volume} {261}},\ \bibinfo {pages}
  {133} (\bibinfo {year} {2016})}\BibitemShut {NoStop}%
\bibitem [{\citenamefont {Liang}\ \emph {et~al.}(2022)\citenamefont {Liang},
  \citenamefont {Ampuero},\ and\ \citenamefont
  {Pino~Mu{\~n}oz}}]{liang2022paucity}%
  \BibitemOpen
  \bibfield  {author} {\bibinfo {author} {\bibfnamefont {C.}~\bibnamefont
  {Liang}}, \bibinfo {author} {\bibfnamefont {J.-P.}\ \bibnamefont {Ampuero}},\
  and\ \bibinfo {author} {\bibfnamefont {D.}~\bibnamefont {Pino~Mu{\~n}oz}},\
  }\bibfield  {title} {\bibinfo {title} {The paucity of supershear earthquakes
  on large faults governed by rate and state friction},\ }\href
  {https://doi.org/10.1029/2022GL099749} {\bibfield  {journal} {\bibinfo
  {journal} {Geophysical Research Letters}\ }\textbf {\bibinfo {volume} {49}},\
  \bibinfo {pages} {e2022GL099749} (\bibinfo {year} {2022})}\BibitemShut
  {NoStop}%
\bibitem [{\citenamefont {Chounet}\ \emph {et~al.}(2018)\citenamefont
  {Chounet}, \citenamefont {Vall{\'e}e}, \citenamefont {Causse},\ and\
  \citenamefont {Courboulex}}]{chounet2018global}%
  \BibitemOpen
  \bibfield  {author} {\bibinfo {author} {\bibfnamefont {A.}~\bibnamefont
  {Chounet}}, \bibinfo {author} {\bibfnamefont {M.}~\bibnamefont {Vall{\'e}e}},
  \bibinfo {author} {\bibfnamefont {M.}~\bibnamefont {Causse}},\ and\ \bibinfo
  {author} {\bibfnamefont {F.}~\bibnamefont {Courboulex}},\ }\bibfield  {title}
  {\bibinfo {title} {Global catalog of earthquake rupture velocities shows
  anticorrelation between stress drop and rupture velocity},\ }\href
  {https://www.sciencedirect.com/science/article/pii/S0040195117304572}
  {\bibfield  {journal} {\bibinfo  {journal} {Tectonophysics}\ }\textbf
  {\bibinfo {volume} {733}},\ \bibinfo {pages} {148} (\bibinfo {year}
  {2018})}\BibitemShut {NoStop}%
\bibitem [{\citenamefont {Bao}\ \emph {et~al.}(2019)\citenamefont {Bao},
  \citenamefont {Ampuero}, \citenamefont {Meng}, \citenamefont {Fielding},
  \citenamefont {Liang}, \citenamefont {Milliner}, \citenamefont {Feng},\ and\
  \citenamefont {Huang}}]{bao2019early}%
  \BibitemOpen
  \bibfield  {author} {\bibinfo {author} {\bibfnamefont {H.}~\bibnamefont
  {Bao}}, \bibinfo {author} {\bibfnamefont {J.-P.}\ \bibnamefont {Ampuero}},
  \bibinfo {author} {\bibfnamefont {L.}~\bibnamefont {Meng}}, \bibinfo {author}
  {\bibfnamefont {E.~J.}\ \bibnamefont {Fielding}}, \bibinfo {author}
  {\bibfnamefont {C.}~\bibnamefont {Liang}}, \bibinfo {author} {\bibfnamefont
  {C.~W.}\ \bibnamefont {Milliner}}, \bibinfo {author} {\bibfnamefont
  {T.}~\bibnamefont {Feng}},\ and\ \bibinfo {author} {\bibfnamefont
  {H.}~\bibnamefont {Huang}},\ }\bibfield  {title} {\bibinfo {title} {{Early
  and persistent supershear rupture of the 2018 magnitude 7.5 Palu
  earthquake}},\ }\href {https://www.nature.com/articles/s41561-018-0297-z}
  {\bibfield  {journal} {\bibinfo  {journal} {Nature Geoscience}\ }\textbf
  {\bibinfo {volume} {12}},\ \bibinfo {pages} {200} (\bibinfo {year}
  {2019})}\BibitemShut {NoStop}%
\bibitem [{\citenamefont {Irwin}(1957)}]{Irwin1957}%
  \BibitemOpen
  \bibfield  {author} {\bibinfo {author} {\bibfnamefont {G.~R.}\ \bibnamefont
  {Irwin}},\ }\bibfield  {title} {\bibinfo {title} {{Analysis of stresses and
  strains near the end of a crack traversing a plate}},\ }\href
  {https://doi.org/10.1115/1.4011547} {\bibfield  {journal} {\bibinfo
  {journal} {J. Appl. Mech.}\ }\textbf {\bibinfo {volume} {24}},\ \bibinfo
  {pages} {361} (\bibinfo {year} {1957})}\BibitemShut {NoStop}%
\bibitem [{\citenamefont {Das}(1980)}]{das1980numerical}%
  \BibitemOpen
  \bibfield  {author} {\bibinfo {author} {\bibfnamefont {S.}~\bibnamefont
  {Das}},\ }\bibfield  {title} {\bibinfo {title} {A numerical method for
  determination of source time functions for general three-dimensional rupture
  propagation},\ }\href {https://doi.org/10.1111/j.1365-246X.1980.tb02593.x}
  {\bibfield  {journal} {\bibinfo  {journal} {Geophysical Journal
  International}\ }\textbf {\bibinfo {volume} {62}},\ \bibinfo {pages} {591}
  (\bibinfo {year} {1980})}\BibitemShut {NoStop}%
\bibitem [{\citenamefont {Breitenfeld}\ and\ \citenamefont
  {Geubelle}(1998)}]{Breitenfeld1998}%
  \BibitemOpen
  \bibfield  {author} {\bibinfo {author} {\bibfnamefont {M.~S.}\ \bibnamefont
  {Breitenfeld}}\ and\ \bibinfo {author} {\bibfnamefont {P.~H.}\ \bibnamefont
  {Geubelle}},\ }\bibfield  {title} {\bibinfo {title} {{Numerical analysis of
  dynamic debonding under 2D in-plane and 3D loading}},\ }\href
  {https://doi.org/10.1023/A:1007535703095} {\bibfield  {journal} {\bibinfo
  {journal} {Int. J. Fract.}\ }\textbf {\bibinfo {volume} {93}},\ \bibinfo
  {pages} {13} (\bibinfo {year} {1998})}\BibitemShut {NoStop}%
\bibitem [{\citenamefont {Brener}\ \emph {et~al.}(2018)\citenamefont {Brener},
  \citenamefont {Aldam}, \citenamefont {Barras}, \citenamefont {Molinari},\
  and\ \citenamefont {Bouchbinder}}]{Brener2018}%
  \BibitemOpen
  \bibfield  {author} {\bibinfo {author} {\bibfnamefont {E.~A.}\ \bibnamefont
  {Brener}}, \bibinfo {author} {\bibfnamefont {M.}~\bibnamefont {Aldam}},
  \bibinfo {author} {\bibfnamefont {F.}~\bibnamefont {Barras}}, \bibinfo
  {author} {\bibfnamefont {J.-F.}\ \bibnamefont {Molinari}},\ and\ \bibinfo
  {author} {\bibfnamefont {E.}~\bibnamefont {Bouchbinder}},\ }\bibfield
  {title} {\bibinfo {title} {{Unstable Slip Pulses and Earthquake Nucleation as
  a Nonequilibrium First-Order Phase Transition}},\ }\href
  {https://doi.org/10.1103/PhysRevLett.121.234302} {\bibfield  {journal}
  {\bibinfo  {journal} {Phys. Rev. Lett.}\ }\textbf {\bibinfo {volume} {121}},\
  \bibinfo {pages} {234302} (\bibinfo {year} {2018})}\BibitemShut {NoStop}%
\end{thebibliography}

%

\end{document}